\documentclass[
reprint,
superscriptaddress,
 amsmath,amssymb,
 aps, physrev,
]{revtex4-2}

\usepackage{graphicx}% Include figure files
\usepackage{dcolumn}% Align table columns on decimal point
\usepackage{bm}% bold math
\usepackage{physics}
\usepackage{xcolor}

\begin{document}

\preprint{APS/123-QED}

\title{\textbf{{Robust multi-mode superconducting circuit optimized for quantum information processing}} 
}%

\author{Pablo García-Azorín}
\affiliation{%
 Department of Physical Chemistry, University of the Basque Country UPV/EHU, Apartado 644, 48080 Bilbao, Spain
}%
\affiliation{%
 EHU Quantum Center, University of the Basque Country UPV/EHU, Apartado 644, 48080 Bilbao, Spain
}%

\author{Francisco A. Cárdenas-López}
\affiliation{%
 Forschungszentrum Jülich GmbH, Peter Grünberg Institute, Quantum Control (PGI-8), 52425 Jülich, Germany
}%

\author{Gerhard B. P. Huber}
\affiliation{%
 Walther-Meißner-Institut, Bayerische Akademie der Wissenschaften, 85748 Garching, Germany
}%
\affiliation{%
 Technical University of Munich, TUM School of Natural Sciences, Department of Physics, Garching 85748, Germany
}%

\author{Guillermo Romero}
\affiliation{%
 Departamento de Física, CEDENNA, Universidad de Santiago de Chile, Avenida Víctor Jara 3493, 9170124, Santiago, Chile
}%

\author{Max Werninghaus}
\affiliation{%
 Walther-Meißner-Institut, Bayerische Akademie der Wissenschaften, 85748 Garching, Germany
}%
\affiliation{%
 Technical University of Munich, TUM School of Natural Sciences, Department of Physics, Garching 85748, Germany
}%

\author{Felix Motzoi}
\affiliation{%
 Forschungszentrum Jülich GmbH, Peter Grünberg Institute, Quantum Control (PGI-8), 52425 Jülich, Germany
}%
\affiliation{%
 Institute for Theoretical Physics, University of Cologne, 50937 Cologne, Germany
}%

\author{Stefan Filipp}
\affiliation{%
 Walther-Meißner-Institut, Bayerische Akademie der Wissenschaften, 85748 Garching, Germany
}%
\affiliation{%
 Technical University of Munich, TUM School of Natural Sciences, Department of Physics, Garching 85748, Germany
}%
\affiliation{%
 Munich Center for Quantum Science and Technology (MCQST), Schellingstraße 4, 80799 München, Germany
}%

\author{Mikel Sanz}
\affiliation{%
 Department of Physical Chemistry, University of the Basque Country UPV/EHU, Apartado 644, 48080 Bilbao, Spain
}%
\affiliation{%
 Basque Center for Applied Mathematics (BCAM), Alameda de Mazarredo, 14, 48009 Bilbao, Spain
}%
\affiliation{%
 EHU Quantum Center, University of the Basque Country UPV/EHU, Apartado 644, 48080 Bilbao, Spain
}%
\affiliation{%
 IKERBASQUE, Basque Foundation for Science, Plaza Euskadi 5, 48009, Bilbao, Spain
}

\date{\today}% It is always \today, today,

\begin{abstract}
{Multi-mode superconducting circuits offer a promising platform for engineering robust systems for quantum computation. Previous studies indicate that single-mode devices cannot be engineered to simultaneously exhibit resilience against multiple decoherence sources due to conflicting requirements. In contrast, multi-mode systems offer increased flexibility and have proven capable of overcoming these fundamental limitations. Here, we present a multi-mode device optimized for quantum information processing. It features an anharmonicity of a third of the qubit frequency and reduced energy dispersion caused by charge and magnetic flux fluctuations. It exhibits improvements over the fundamental errors limiting Transmon and Fluxonium coherence and control, achieving ratios between the total coherence time and the gate time $T_2/t_g$ one order of magnitude larger than Transmon and two times larger than Fluxonium for microwave charge drives, assuming equal dielectric and inductive loss quality factors and limited drive strength. It furthermore demonstrates robustness against fabrication errors, a major limitation in many proposed multi-mode devices.}
\end{abstract}

\maketitle

%\tableofcontents

\section{\label{sec:introduction}Introduction}
Among the various platforms for realizing quantum information processing, superconducting qubits have emerged as one of the leading contenders due to their scalability, controllability, and compatibility with existing semiconductor fabrication techniques. However, the widespread adoption of superconducting qubits for practical quantum computation faces significant challenges, primarily stemming from its scalability and limited coherence times.

Since the first proposals of superconducting systems to perform quantum computation, {their} capabilities and computational power have been limited by decoherence. Errors affecting superconducting devices have been traditionally classified into two main categories: errors producing dephasing and errors producing depolarization. In the development of the first superconducting qubits, starting from the Cooper-Pair Box \cite{Bouchiat1998, Nakamura1999} and Flux qubit \cite{Mooij1999}, dephasing was the limiting factor constraining operability due to the large sensitivity to environmental fluctuations. To overcome the initial limitations, superconducting systems were engineered to suppress specific noise vulnerabilities, a {representative} example of that being the Transmon device \cite{PhysRevA76042319}, {which exhibits} an exponentially suppressed sensitivity to external charge bias. Despite this improvement, the Transmon still suffers {from} some limitations, mainly stemming from the reduced anharmonicity and large dipole matrix elements that lead to a strong dissipation and induce depolarization. Strategies to mitigate \cite{Paik2011} and to model \cite{Ganjam2024} {this phenomenon} have been widely studied, leading to advancements in coherence \cite{Place2021,Wang2022}. {At the same time, an alternative approach to addressing charge sensitivity was proposed in the form of the Fluxonium device~\cite{Manucharyan2009}, which incorporates an additional shunting inductor to reduce charge matrix elements, at the expense of increased flux sensitivity.} By relying on the reduced amplitude of external magnetic flux fluctuations compared to charge fluctuations, coherence times were improved. Further advancements in quasiparticle suppression \cite{Nature508}, improvement over dielectric losses at low frequencies \cite{PhysRevX9041041}, and the exploration of very-low frequency regimes \cite{PhysRevX11011010}, led to a significant improvement in coherence times \cite{Somoroff2023}. {However, beyond these limitations, practical operation requires the system to exhibit large anharmonicity and large charge (flux) matrix elements in order to enable fast gate operations and coupling to neighboring qubits.} This further imposes another tradeoff {between} matrix elements, anharmonicities and coherence times. {Consequently, fundamentally different error sources require specific system characteristics that may conflict in certain scenarios, thereby complicating the realization of fully robust systems}~\cite{Gyenis2021}.

Most of the devices mentioned above rely on a single degree of freedom to encode the quantum information, that is, one single mode, which can be physically understood as the number of Cooper pairs contained on an island of superconducting material, the superconducting phase difference between two of them or even a persistent current oscillating in a superconducting loop. In recent years, the difficulties in finding proper tradeoffs for single-mode devices have motivated the exploration of more complex circuits, with an increasing number of elements constituting multi-mode systems \cite{Calzona2023}. Most of the multi-mode device proposals made so far rely on the construction of a protected subspace \cite{Ioffe2002,Gladchenko2009,Douçot2012}, creating an encoding of the computational space in degenerate states expressing exponentially suppressed tunneling amplitudes and external effects sensitivity \cite{Brooks2013,Smith2020,Kalashnikov2020}. However, this still imposes significant challenges for {control, requiring operations that take the system out of the protected state and may still result in decoherence} \cite{Brooks2013,Gyenis2021}. A different strategy is to consider multi-mode architectures without fundamental full state protection, but with improved tradeoffs between noise protection and controllability \cite{Hyyppä2022,Mencia2024}. Nevertheless, the design of systems with an increasing number of elements carries major difficulties. {Not only does the complexity of exploring the configuration space scale dramatically with the number of free parameters, but the analysis of the properties of each configuration also scales exponentially with the number of modes. This fact, combined with the difficulty of forming physical intuition for these complex systems, renders the engineering and optimization of multi-mode configurations a challenging task.}

In this article, {we introduce a multimode device and identify a specific parameter regime for its operation, which we refer to as the Difluxmon qubit, which exhibits characteristics suitable for quantum computation.} The system was derived using computer-aided optimization techniques, in the form of evolutionary algorithms, based on the techniques developed in \cite{Cardenas2023}, to efficiently explore the large parameter space, aiming to minimize the number of iterations and computational power needed. {At the optimized parameter regime and operation point, }the system presents an anharmonicity {of} $\alpha \approx 2\pi\times750$ MHz at a qubit frequency {of} $\omega_{10}=2\pi\times2.5~$GHz, and charge operator matrix elements {of} $|\langle 1 | \hat{n} | 0 \rangle | \approx 0.4$ for microwave control. {It features sufficiently large anharmonicity and qubit frequency to enable fast drive operations, along with a charge dipole moment compatible with the limited drive strength of the control signals. Excessive drive amplitudes can introduce incoherent errors and heating, thereby limiting fast operation in low–dipole-moment qubits~\cite{Rower2024}}. We estimate $T_2 \sim 2T_1 \sim 400~\mu\text{s}$ coherence times, having dielectric losses as the limiting factor. Transition rules were designed to overcome the most common leakage channels, canceling the $|1\rangle \rightarrow |2\rangle $ transition. The reduction of energy dispersion due to charge and flux {noise} was targeted, with the goal of increasing dephasing times. By finding proper matrix element values, allowing for the implementation of fast operations while maintaining certain depolarization and dephasing resilience, {together with an} increased qubit frequency and anharmonicity, we {obtain} a system {with a large ratio between total coherence time and gate time}.

\section{\label{sec:superconducting circuit}Superconducting Circuit Proposal}
\begin{figure*}
\includegraphics[width=\linewidth]{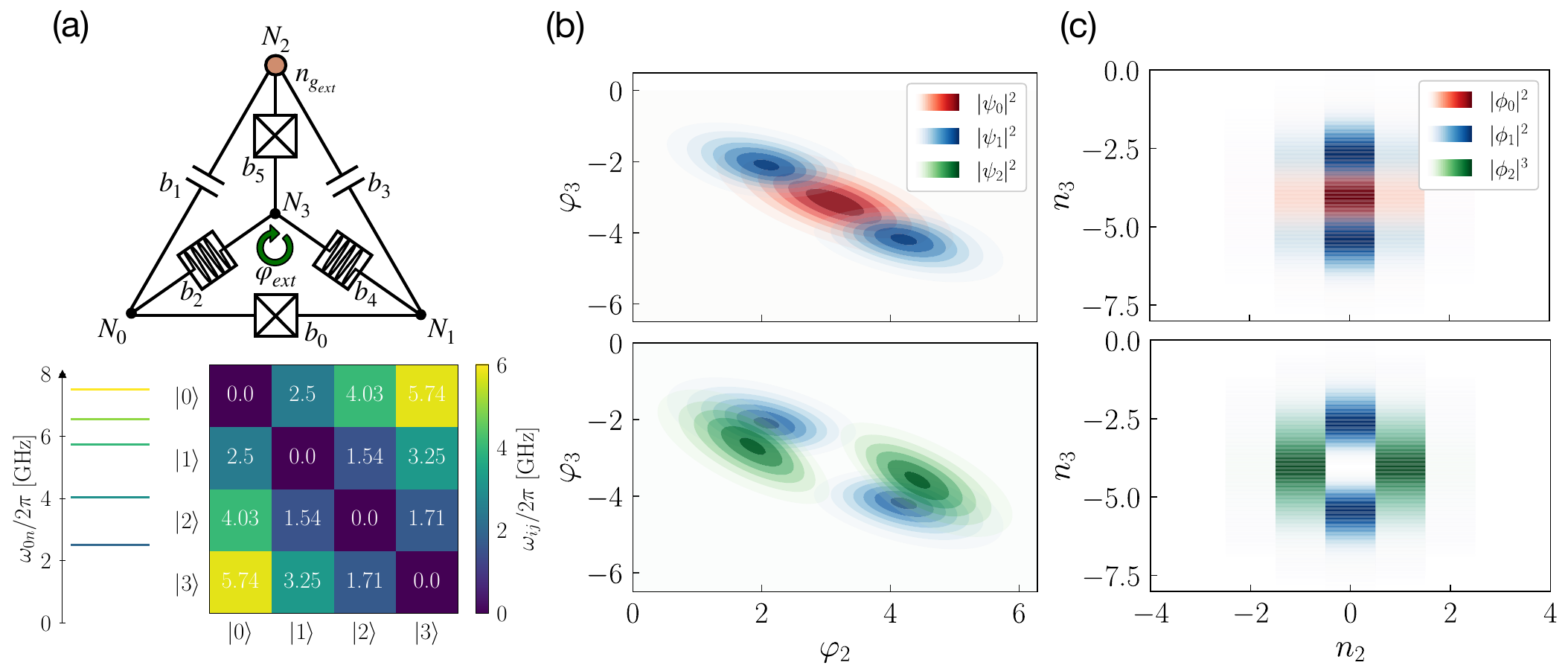}
\caption{\label{fig:schemspect}(a) Lumped-element design consisting of four nodes  $\mathcal{N} = \{N_{0}, N_{1}, N_{2}, N_{3}\}$, and five branches $\mathcal{B} = \{b_{0}, b_{1}, b_{2}, b_{3}, b_{4}\}$, where boxed elements denote the presence of an additional capacitor in parallel; and global energy spectrum of the system for all three modes at $\varphi_{ext} = \pi$, where each element of the matrix represents the energy difference between states $|i\rangle$ and $|j\rangle$. (b) Two-dimensional phase space projection of the probability density onto flux space, (c) and charge space, tracing out spatial direction $1$ for the first three eigenstates.}
\end{figure*}
The circuit {consists of} four nodes in two superconducting islands. The first island {is composed of} two inductors connected in series and shunted by a Josephson junction, forming a superconducting loop. The second island is coupled to this loop via an additional Josephson junction, resulting in a three-mode system. Additionally, it presents capacitive connections between every node in the circuit. The lumped-element schematic of the system is shown in Fig.~\ref{fig:schemspect}(a), where boxed elements denote the presence of capacitances in parallel. {This constitutes a general class of multi-mode circuit~\cite{weissler2024}, for which we define specific parameter values and operation point, giving rise to the Difluxmon qubit.} The circuit parameters considered are shown in Table \ref{table:param}. The system was derived from the optimization of a general four node structure, considering capacitive connections between every node, to mimic the realistic capacitive relations in 2D lithographic designs, and either a linear or non-linear inductor connection (see Appendix \ref{sec:appendixD} for details).
\subsection{Hamiltonian}
The Hamiltonian modeling the device, in terms of node variables (see Appendix \ref{sec:appendixA}), is given by
\begin{eqnarray}\label{eq:Hamiltonian}
\hat{H} &=& \, 4\, \boldsymbol{n}^T \boldsymbol{E}_C\, \boldsymbol{n} + \frac{1}{2}\, \boldsymbol{\varphi}^T \boldsymbol{E}_L\, \boldsymbol{\varphi}\nonumber\\
 & &- E_J^{b_0} \cos{\left( \hat{\varphi}_1\right)} - E_J^{b_5} \cos{\left( \hat{\varphi}_2 - \hat{\varphi}_3\right)}\\
 & &+ \hat{H}_{\text{ext}}\!\left(n_{g_{\text{ext}}}, \varphi_{\text{ext}}\right),\nonumber
\end{eqnarray}
where $\boldsymbol{n}=(\hat{n}_1,\hat{n}_2,\hat{n}_3)^T$ and $\boldsymbol{\varphi}=(\hat{\varphi}_1,\hat{\varphi}_2,\hat{\varphi}_3)^T$ are the charge and phase vectors, choosing node $N_0$ as reference; with $\hat{n}_i=\hat{q}_i/(2e)$ and $\hat{\varphi}_i=(2\pi/\Phi_0)\hat{\Phi}_i$, being $\hat{q}_i$ and $\hat{\Phi}_i$ the charge and flux operators of node $i$ respectively; ${[E_C]}_{ij} = (e^2/2){[C^{-1}]}_{ij}$ and ${[E_L]}_{ij} = \left(\Phi_0/2\pi\right)^2{[L^{-1}]}_{ij}$ represent the matrices of charging and inductive energies respectively, with $\Phi_0$ being the flux quantum and $e$ the electron charge. From the circuit, we observe a system of three modes, one \textit{periodic} \{$N_2$\} and two \textit{extended} $\{N_1, N_3\}$, strongly coupled together. The system is subjected to contributions from external charge bias $n_{g_{\rm ext}}$ and external magnetic flux bias $\varphi_{\rm ext}$, modeled by $\hat{H}_{\text{ext}}$. The effect of these external factors will be studied later in the article.
\subsection{Energy Spectrum} \label{subsec:Energy Spectrum}
The numerically computed energy spectrum of the device at the operation point $\varphi_{\text{ext}} = 0.5$ is shown in Fig.~\ref{fig:schemspect}(a)  (see Appendix \ref{sec:appendixB} for details on the numerical representation). We observe a highly anharmonic energy spectrum; the computational states present a frequency difference $\omega_{10}=2\pi\times 2.5~$ GHz, above thermal frequency assuming operation temperatures of $15-20$ mK ($\sim 0.4$ GHz). It presents an anharmonicity $\eta=\omega_2-2\omega_1=2\pi\times 1~$ GHz with respect to the $\ket{1}\leftrightarrow\ket{2}$ transition. However, this leakage channel is further suppressed {as a result of the cancellation of the matrix elements coupling} $\ket{1}$ and $\ket{2}$ states ($|\langle 2 | \hat{n}_i | 1 \rangle| \approx 0$). Instead, the main leakage channel corresponds to the $\ket{1}\leftrightarrow\ket{3}$ transition that has a positive anharmonicity $\alpha=\omega_3-2\omega_1\approx2\pi\times 750~$ MHz.
\subsection{Matrix Elements} \label{subsec:Matrix Elements}
\begin{figure}[b]
\includegraphics[width=1\linewidth]{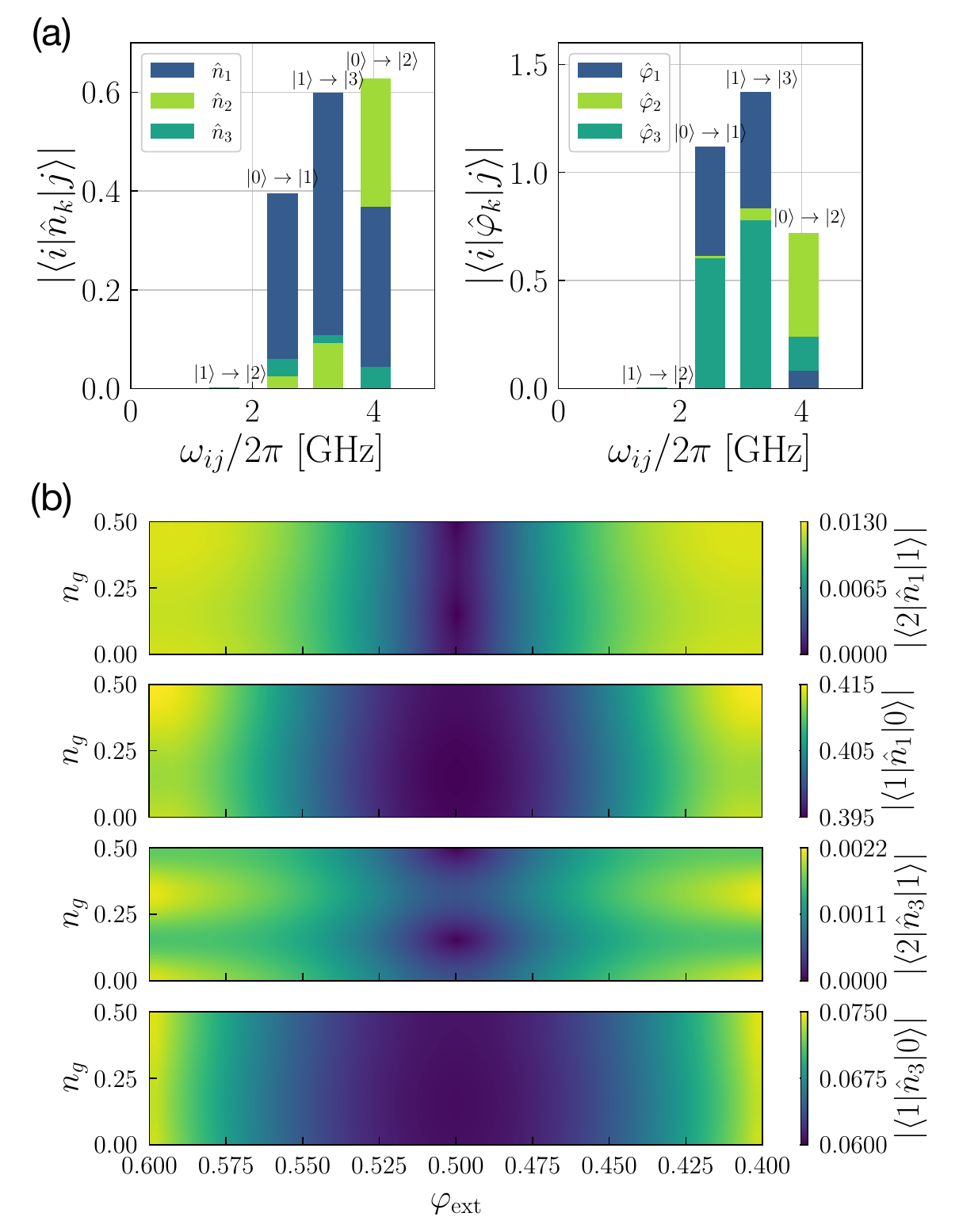}
\caption{\label{fig:matelem}(a) Charge and flux-node operators matrix elements for all three nodes in the system $\{ N_1, N_2, N_3\}$ with respect to the transition frequencies of the first three eingestates of the device at the operation point $\varphi_{\text{ext}} = 0.5$. {(b) Charge matrix elements of $N_1$ and $N_3$, of interest for driving and measurement purposes, with respect to the external magnetic flux and charge bias.}}
\end{figure}

The matrix elements of charge and flux node operators dictate the coupling strength to the {electromagnetic} environment and the control lines. Consequently, decreasing the coupling strength intuitively increases the coherence time. However, at the same time, high isolation from the {electromagnetic environment} can make the system uncontrollable and difficult to measure. For that reason, it is necessary to design devices with a balance between noise protection and control. The node charge and flux matrix elements of the device are shown in Fig.~\ref{fig:matelem}(a). Focusing on the local charge matrix elements, if we select node $N_1$ as our coupling point for driving purposes, we observe a decrease of over a factor of 2 from the usual charge matrix element values in Transmon devices ($|\langle \hat{n} \rangle | \sim 1$) \cite{PhysRevA76042319}. Despite this reduction, the values are still sufficiently large for external driving through a capacitively coupled microwave line, overcoming the controllability problems produced from the strong reduction of charge matrix elements in devices such as Fluxoniums ($|\langle \hat{n} \rangle | \sim 0.1)$) \cite{PhysRevX9041041, PRXQuantum3037001}, Heavy Fluxoniums ($|\langle \hat{n} \rangle | \sim 0.01$) \cite{PhysRevX11011010} or parity protected devices \cite{Smith2020, PRXQuantum2010339}, and allowing for the implementation of optimal control techniques developed for charge driven systems. Furthermore, we observe protection of the computational space from charge noise fluctuations affecting all other islands in the system, resulting from the reduction of the matrix elements for the two nearest transitions affecting the computational states ($|1\rangle \rightarrow |2\rangle$ and $|1\rangle \rightarrow |3\rangle$) for the rest of the circuit nodes $\{ N_2, N_3 \}$ (see Fig.~\ref{fig:nphiop_mat}(a)). The magnitude of flux matrix elements is also suppressed compared to implemented Fluxonium and Heavy Fluxonium devices ($|\langle \hat{\varphi} \rangle | \sim 1.5-2)$) \cite{PhysRevX9041041,PhysRevX11011010}, and due to the fact that flux fluctuations are typically around two orders of magnitude smaller than charge fluctuations in experimental setups \cite{Gyenis2021}, we will not expect them to represent a limiting factor.
\subsection{External Biases} \label{subsec:External Biases}
\begin{figure*}
\includegraphics[width=0.85\linewidth]{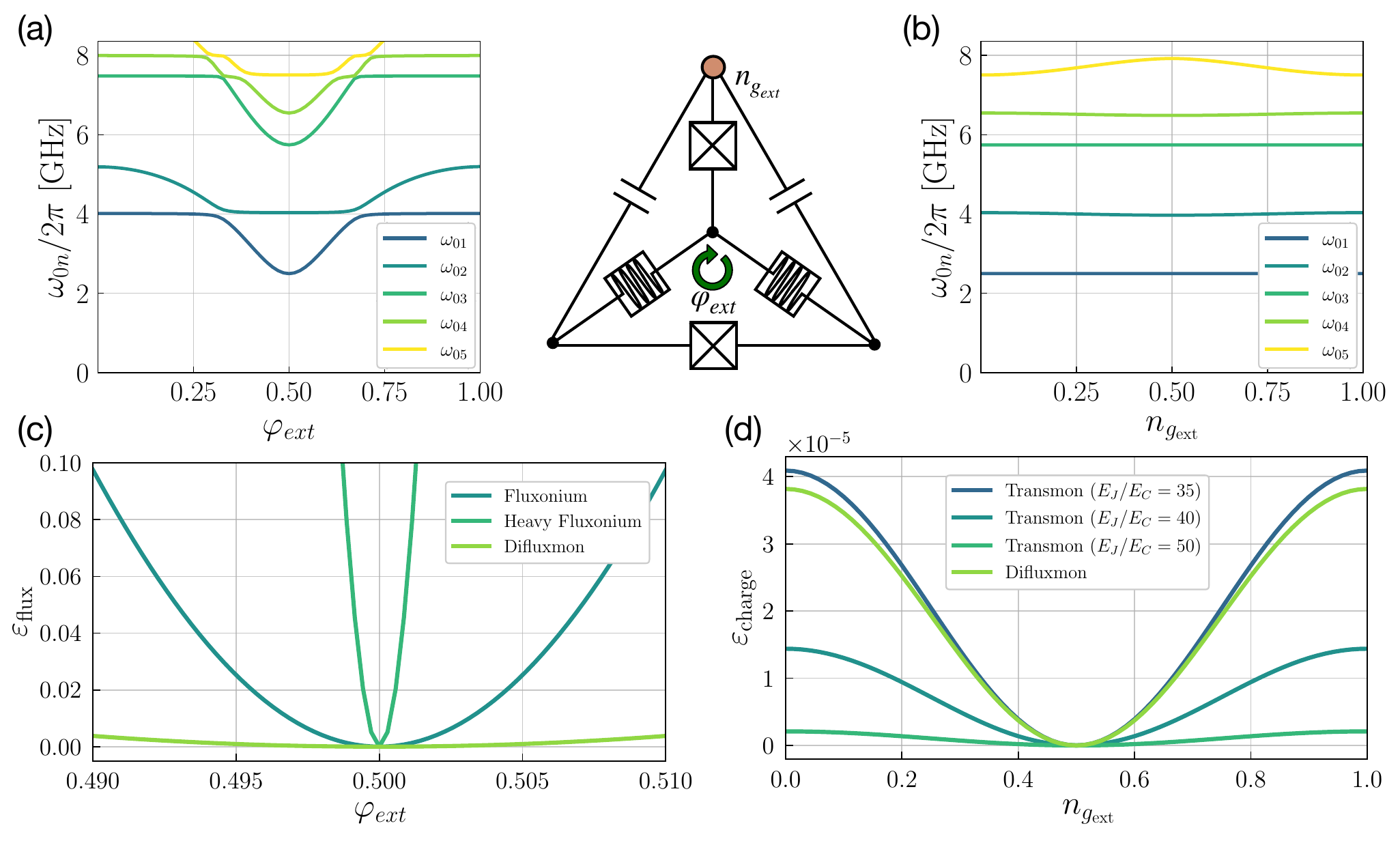}
\caption{\label{fig:nphiop_mat}(a, b) Effect of external flux $\varphi_{\rm ext}$ and charge $n_{g_{\rm ext}}$ bias, respectively, on the energy spectrum of the system. In ascending order, the energy difference $\omega_{0n} = \omega_n - \omega_0$ is shown for $n \in \{1,2,3,4,5\}$ for different values of external magnetic flux (charge) bias. (c, d) Energy dispersion with respect to external effects. (c) Plot of the energy dispersion $\varepsilon_{\text{flux}} = \left(\omega_{01} - \omega_{01}|_{\varphi_{\text{ext}}=0.5}\right)/\left(\omega_{01}|_{\varphi_{\text{ext}}=0.5}\right)$ for different values of external flux bias around the operation point for the optimized device (OPT), in comparison with proposed Fluxonium \cite{PRXQuantum3037001} and Heavy Fluxonium \cite{PhysRevX11011010} devices. (d) Plot of the energy dispersion $\varepsilon_{n_{g_{\text{ext}}}} = \left(\omega_{01} - \omega_{01}|_{n_{g_{\text{ext}}}=0.5}\right)/\left(\omega_{01}|_{n_{g_{\text{ext}}}=0.5}\right)$ for different values of external charge bias, in comparison with different Transmon regimes.}
\end{figure*}

In contrast to typical single-mode devices, the proposed system is susceptible to both external charge and flux bias. The susceptibility to external charge bias is due to the presence of a \textit{periodic} mode, or equivalently, a superconducting island coupled to the rest of the system only by capacitors and Josephson junctions, located in node $N_2$ of the circuit. The susceptibility to external flux is due to the presence of a closed loop formed by inductors and Josephson junctions, observed in the series connection of the branches $\{b_0, b_4, b_2\}$ (see Fig. \ref{fig:schemspect}(a)). The addition of the aforementioned effects can be modeled including
\begin{eqnarray}\label{eq:extHamiltonian}
\hat{H}_{\text{ext}} &=&\, 8\left( [E_C]_{11}\hat{n}_2 + [E_C]_{01}\hat{n}_1 + [E_C]_{12}\hat{n}_3 \right) n_{g_{\text{ext}}}\\
& &+ [E_L]_{b_2}\hat{\varphi}_3\,\varphi_{\text{ext}}\nonumber
\end{eqnarray}
to the description, where $\{b_2\}$ was chosen as the closure branch of the inductive loop, $[E_C]_{ij}$ are the charging energies defined by the charge energy matrix and $[E_L]_{b_2} = \frac{1}{L_{b_2}}\left(\frac{\Phi_0}{2\pi}\right)^2$, with $\Phi_0$ being the flux quantum. The effect of these external biases on the energy eigenvalues of the system is shown in Fig.~\ref{fig:nphiop_mat}(a,b). Focusing on the effect of external flux bias, we observe a reduced dispersion around the operation point $\varphi_{\text{ext}} = 0.5$. For fluctuations in external flux of around $10^{-2} \Phi_0$, we can observe a dispersion of less than 10 MHz. For comparison, in Fig. \ref{fig:nphiop_mat}(c) we represent the energy dispersion from the operation point for several Fluxonium devices. We can observe that the proposed device expresses a dispersion of less than $0.5\%$ from the qubit frequency, considerably small compared to other flux-sensitive devices \cite{PRXQuantum3037001, PhysRevX11011010, PhysRevX9041041}. 
When considering the effect of external charge bias, we observe that the dispersion is around 90 kHz, which is in the order of state-of-the-art charge-sensitive devices employed for quantum computation. In Fig. \ref{fig:nphiop_mat}(d) we computed the energy dispersion due to external charge bias effect and checked that it accounts for a dispersion of $0.004\%$ of the operation frequency. Furthermore, we compare with the Transmon device in different parameter regimes, showing dispersions in the order of the proposed device.
For completeness, we further present the dispersion of the charge operator matrix elements for nodes $N_1$ and $N_3$ in Fig. \ref{fig:matelem}(b), showing a reduced sensitivity around the operation point $\varphi_{\text{ext}}=\pi$, suitable for gate operation and readout.
\begin{table}[b]%The best place to locate the table environment is directly after its first reference in text
\caption{\label{tab:table1}%
Circuit parameter of the device depicted in Fig.~\ref{fig:schemspect}(a).
}
\begin{ruledtabular}\label{table:param}
\begin{tabular}{lcdr}
\multicolumn{4}{c}{Circuit parameters}\\
\colrule
Branch    & $C~$(fF) & $L~$(nH) & $E_{J}~$(GHz)  \\
\colrule
$b_0$     & 11.62 & -- & 2.5  \\
$b_1$       & 12.48 & -- & -- \\
$b_2$   & 15.31 & 35.21 & -- \\
$b_3$        & 12.29 & -- & -- \\
$b_4$   & 10.27 & 32.82 & -- \\
$b_5$   & 10.94 & -- & 6.85\\
\end{tabular}
\end{ruledtabular}
\end{table}
\subsection{Wave functions}
To gain a deeper insight into the system's characteristics, it is beneficial to analyze its wave functions. Considering the three-dimensional potential landscape of the device for the parameters listed in Table~\ref{table:param}, the wave functions of the first three eigenstates remain localized within the same potential well. However, an analysis of their phase-space representation reveals that the ground, first, and second excited states exhibit reduced overlap with one another. This limited spatial support mitigates external depolarization effects and further contributes to the suppression of magnetic flux dephasing, which couples to the system through the $\varphi_3$ operator. Fig.~\ref{fig:schemspect}(b) presents two-dimensional phase-space projections for modes $\{2,3\}$ in flux space (see Appendix~\ref{sec:appendixW}). Additionally, the system’s resilience to charge fluctuations is illustrated through the charge-space projections in Fig.~\ref{fig:schemspect}(c). In our model, charge fluctuations couple to the system through node two, as it represents the only periodic mode in the device. For the computational basis states, encoded in the first two wave functions, fluctuations along $n_2$ direction do not induce transitions, as they satisfy $\langle\phi_0|\hat{n}_2|\phi_0\rangle \approx \langle\phi_1|\hat{n}_2|\phi_1\rangle \approx 0$.
Transitions to the third eigenstate $\phi_2$ are theoretically possible but are strongly suppressed due to the large energy separation. Finally, matrix elements cancellation between the first and second excited states can be understood due to the wave functions parity. Both ground $\hat{U}_P|\psi_0\rangle = |\psi_0\rangle$ and first excited $\hat{U}_P|\psi_1\rangle = -|\psi_1\rangle$ states present positive and negative parity, respectively. However, the second excited state $\hat{U}_P|\psi_2\rangle = -|\psi_2\rangle$ presents also negative parity, consequently for any local operator $\hat{O}$ (i.e. $\hat{n}_i$ and $\hat{\varphi}_i$) such that $\hat{U}_P^\dagger\hat{O}\hat{U}_P = -\hat{O}$ we have $\langle\psi_2|\hat{O}|\psi_1\rangle = \langle\psi_2|\hat{U}_P^\dagger\hat{O}\hat{U}_P|\psi_1\rangle = -\langle\psi_2|\hat{O}|\psi_1\rangle = 0$ \cite{Egusquiza2022}.
\section{\label{sec:CTE}Coherence time estimation}
In order to perform an estimation of the coherence time of the device, focus has been made on the main decoherence channels considered in literature. In the estimation, different types of errors have been classified into two main categories, distinguishing between noises causing depolarization and dephasing.
\subsection{Depolarization noise}
To model the depolarization of the device due to different noise sources, we have employed the traditional expression derived from Fermi's Golden rule \cite{Leggett1987}. As the device on hand presents multiple elements, creating multiple branches $\mathcal{B}$, to compute the total coherence time we have computed the coherence times of every individual branch and added them together, to get a lower bound estimation \cite{Groszkowski2018,Groszkowski2021}. In general, for every depolarization channel, we will have
\begin{eqnarray}\label{eq:T1}
\Gamma_1^{\lambda} = \sum_{b_n \in \mathcal{B}} \sum_{ij} \frac{1}{\hbar^2} \abs{\langle i | \hat{\mathcal{O}}_{\lambda}^{b_n} | j \rangle}^2 S_{\lambda}^{b_n}(\omega_{ij}),
\end{eqnarray}
where at least $i$ or $j$ correspond to one of the computational states $\{0,1\}$, summing over all possible energy transitions into or out from the computational space, being $\hat{\mathcal{O}}_{\lambda}^{b_n}$ the operator coupling to the noise source and $S_{\lambda}^{b_n}(\omega_{ij})$ the power spectral density of the noise source. The consideration of all possible transitions involving at least one of the computational states, in contrast with other expressions used for single-mode devices, is due to the rich variety of possible transitions in the system, where depolarization due to excitation out of the computational space, although exponentially suppressed by the Boltzmann factor, can become one detrimental factor\cite{Groszkowski2018}. The main depolarization mechanisms considered are dielectric losses, inductive losses, and quasiparticle tunneling.
\subsubsection{Dielectric losses}
The appearance of an electric field across islands in the superconducting circuit causes the polarization of charges in the dielectric material to create electric dipole moments that cause dissipation \cite{Pozar}. To model this kind of loss we are going to consider the charge operator as coupling to the noise source $\hat{\mathcal{O}}_{\lambda} = -2e \hat{n}$ and a power spectral density of the form \cite{Ithier2005,PhysRevX11011010}
\begin{eqnarray}\label{eq:SDiel}
S_\text{Diel}(\omega_{ij}) = \left(\frac{2\hbar}{CQ_{\text{Cap}}}\right) F(\omega_{ij},T),
\end{eqnarray}
where
\begin{eqnarray}
F(\omega_{ij}, T) = \frac{\coth{\left(\frac{\hbar|\omega_{ij}|}{2K_bT}\right)}}{1 + \exp\left(\frac{-\hbar\omega_{ij}}{K_bT}\right)},
\end{eqnarray}
$C$ represents the capacitor of the branch, $Q_{\text{Cap}}$ is the {effective dielectric} quality factor of the capacitor, $T$ is the assumed temperature of the device and $K_b$ is the Boltzmann constant. The computed $T_1^{\text{Diel}}$ times for different values of external flux are shown in Fig.~\ref{fig:CoherenceTimes}(a). In the same figure, we display a comparison with Transmon device with a $E_J/E_C \sim 50$ relation and $E_C = 0.27~\text{GHz} $, which accounts for a capacitance of $ \sim\!70~\text{fF}$, reasonably large considering usual qubit sizes of $\sim 600-800\ \mu m^2$. We observe that the estimated depolarization time for the device at the operation point $\varphi_{\text{ext}} = 0.5$ doubles the one expected for the Transmon device assuming equal {dielectric loss} quality factors. Furthermore, we also compare with several Fluxonium devices, where the stronger suppression of the charge matrix elements produces an increase in coherence time, overcoming the $T_1^{\rm Diel}$ expressed by the proposed configuration at the cost of a decrease in controllability.
\subsubsection{Inductive losses}
Similarly as for dielectric losses, we can consider the inductors of the system as having a lossy permeability inducing a frequency-dependent resistance \cite{Pozar}. To model this mechanism we considered the phase difference operator in every inductor $\hat{\mathcal{O}}_{\lambda} = \Phi_0\hat{\varphi}/(2\pi)$ and a power spectral density of the form \cite{Ithier2005,PhysRevX11011010}
\begin{eqnarray}
S_\text{Ind}(\omega_{ij}) = \left(\frac{2\hbar}{LQ_{\text{Ind}}}\right) F(\omega_{ij},T),
\end{eqnarray}
where $L$ represents the inductance of the branch and $Q_{\text{Ind}}$ accounts for the {effective} quality factor {associated with inductive losses}. After estimating the effect of this loss mechanism, we concluded that resistive loss in the inductors is unlikely to be the primary factor limiting coherence in the device, with coherence times on the order of milliseconds. This we believe is due to the lack of large {inductances} and consequently, small phase matrix elements, in contrast with other flux-sensitive devices where this noise mechanism is more notable.
\subsubsection{Quasiparticle tunneling}
The tunneling of quasiparticles in the Josephson junction elements is another well-known mechanism of dissipation and decoherence \cite{PhysRevB84064517,Nature508}. To model this effect we consider the operator $\hat{\mathcal{O}}_{\lambda} = 2\Phi_0\sin{\left(\hat{\varphi}/2\right)}$. To obtain a numerical representation of this operator in the charge basis, we performed a redefinition for the "charge-like" degree of freedom of $N_2$, to the charge basis defined by single electrons instead of Cooper pairs \cite{PhysRevB84064517} (see Methods section). Additionally, a noise spectral density of the form
\begin{eqnarray}
S_{\text{qp}}(\omega_{ij}) = 2\hbar\omega_{ij} \Re{Y_{\text{qp}}(\omega_{ij})} F(\omega_{ij},T),
\end{eqnarray}
was considered, where $\Re{Y_{\text{qp}}(\omega_{ij})}$ can be approximated via the expression proposed in \cite{Smith2020}. From numerical simulations, we concluded that quasiparticle tunneling matrix elements vanish at the operation point of $\varphi_{\text{ext}} = 0.5$, similarly to previously reported flux sensitive devices proposed for operation at the sweet spot \cite{Nature508, Smith2020}. Consequently, this depolarization mechanism is not expected to constitute one of the most detrimental factors constraining the coherence times.
\subsection{Dephasing errors}
For the estimation of the dephasing rate of the device, we are going to focus on one of the expected main contributions, which will come from the $1/f$ noise sources. These sources of noise are characterized by a power spectral density \cite{PhysRevX9041041}
\begin{eqnarray}
S_{1/f}^\lambda(\omega) = \frac{2\pi A_\lambda^2}{|\omega|},
\end{eqnarray}
where, away from the sweet spot, they cause a dephasing rate of \cite{PhysRevX11011010, PhysRevX9041041, PhysRevA76042319,Groszkowski2021}
\begin{eqnarray}
\Gamma_{1/f}^\lambda = \sqrt{2} A_\lambda \left|\frac{\partial \omega_{01}}{\partial \lambda}\right|\sqrt{|\ln{\omega_{\text{ir}}t}|}.
\end{eqnarray}
\subsubsection{$1/f$ flux noise}
One of the main sources of decoherence in current flux-sensitive devices is the $1/f$ flux noise coming from the energy dispersion caused by the external magnetic flux $\varphi_{\text{ext}}$. For the estimation of the dephasing rate due to this noise source, we considered a reference value for the noise amplitude $A_{\varphi_{\text{ext}}}$ of $10^{-6} \Phi_0$, around the experimentally observed values \cite{PhysRevX9041041, PhysRevX7031037, PhysRevLett97167001,Mencia2024}, and we analytically computed the derivative of the qubit frequency from the Hamiltonian expression of Eq.~\ref{eq:extHamiltonian} as $\partial \omega_{01}/\partial \Phi_{\text{ext}} = 1/\hbar \left[ \langle 1 | \partial \hat{H}_{\text{ext}}/\partial \Phi_{\text{ext}}|1 \rangle - \langle 0 | \partial \hat{H}_{\text{ext}}/\partial \Phi_{\text{ext}} |0\rangle \right]$ where $\partial \hat{H}_{\text{ext}}/\partial \Phi_{\text{ext}} = [E_L]_{b_2} \left( \hat{\varphi}_3 + \varphi_{\text{ext}}\mathbf{1} \right)$. The estimated dephasing times for different values of external flux are shown in Fig.~\ref{fig:CoherenceTimes}(b). We observe that the reduced energy dispersion causes an improvement in the estimated dephasing time. In the same figure, we compare with the estimated dephasing times for several Fluxonium devices, where this effect is one of the most detrimental factors for the total coherence time.
\subsubsection{$1/f$ charge noise}
Similarly as for the flux mechanism, we considered the energy dispersion due to charge bias effects as a possible mechanism inducing dephasing. Due to the reduced energy dispersion caused by external charge bias (see Fig.~\ref{fig:nphiop_mat}(d)), around $0.005\%$ of the qubit frequency, this constitutes a minor decoherence channel, and after estimation, we observed that the coherence limit established by this mechanism is well above the one dictated from other decoherence channels.
\subsection{$T_2$ times estimation}
\begin{figure*}
\includegraphics[width=\linewidth]{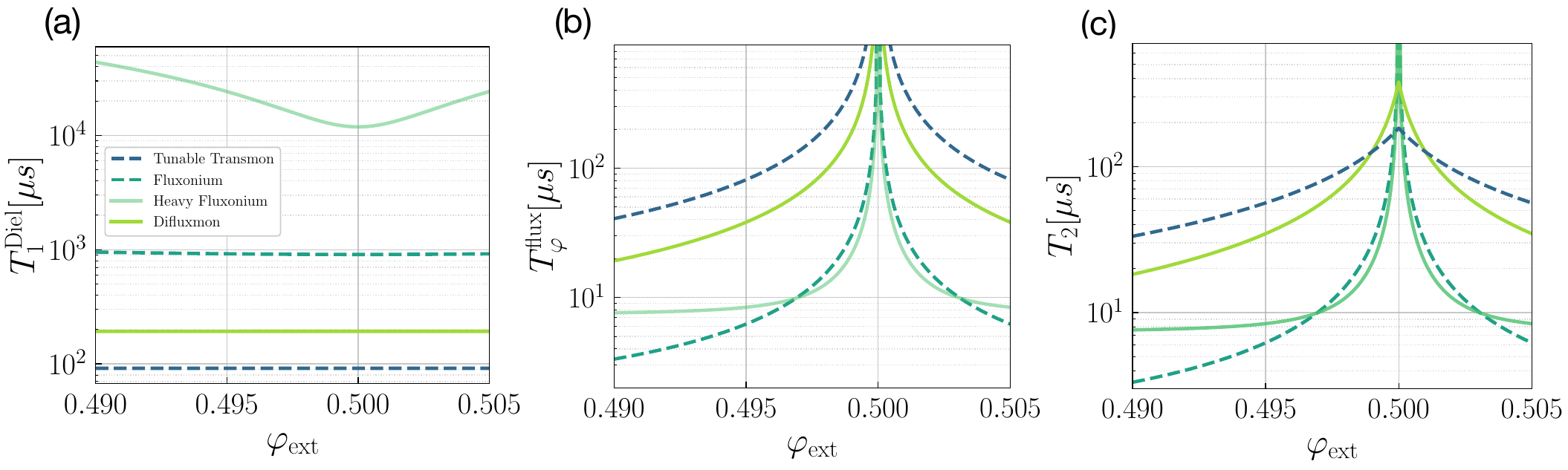}
\caption{\label{fig:CoherenceTimes} Coherence times estimation for the most detrimental decoherence channels. (a) $T_1$ estimation due to dielectric losses assuming $Q_{\text{Cap}} = 3\cdot10^6$ and $T=15$ mK. (b) $T_{1/f}^{\text{Flux}}$ dephasing time estimation due to fluctuations of the external magnetic flux $\varphi_{\text{ext}}$, assuming $A_{\varphi_{\text{ext}}} = 10^{-6} \Phi_0$. (c) $T_2$ times estimation. For comparison, estimated coherence times for the Difluxmon, tunable Transmon $\{ \omega_{01} \approx 5.2 \,\rm{GHz}, E_J/E_C \approx 50, \gamma=2.5\}$~\cite{Krantz2019}, Fluxonium~\cite{PRXQuantum3037001} and Heavy Fluxonium~\cite{PhysRevX11011010} devices are shown assuming equal quality factors and external fluctuation amplitudes.}
\end{figure*}

In order to account for all depolarization and dephasing effects, we consider the usual expression $T_2 = (1/2T_1 + 1/T_{\varphi})^{-1}$, defining the total depolarization time as $T_1 = (1/T_1^{\text{Diel}} + 1/T_1^{\text{Res}} + 1/T_1^{\text{qp}})^{-1}$ and the total dephasing time as $T_{\varphi} = (1/T_{\varphi}^{\text{flux}} + 1/T_{\varphi}^{\text{charge}})^{-1}$. The estimated $T_2$ for different external flux values are shown in Fig.~\ref{fig:CoherenceTimes}(c). We observe that around the operation point $\varphi_{\text{ext}} = 0.5$, dielectric losses limit the coherence time. However, we can estimate improved robustness to magnetic flux fluctuations than Fluxonium devices away from the sweet spot, and larger coherence times than tunable Transmon up to fluctuations of $\sim 10^{-3}\Phi_0$. These coherence characteristics, considering the increase in anharmonicity and charge matrix elements, lead to a device with properties balancing noise resilience and controllability.
\section{Single-qubit gates} \label{sec:sqg}
Having studied the selection rules and coherence properties of our multi-mode artificial atom, the next step is to analyze which are the gates to be implemented in the platform. In our case, in light of the selection rules of the circuit operators provided in Fig.~\ref{fig:matelem}(a), we aim to implement $\{\mathcal{R}_{x}(\theta), \mathcal{R}_{y}(\theta)\}$ gates at shorter time $t_{g}$. Our goal is to analyze the gate fidelity as well as the leakage channels for different gate times. We assume charge-controlled quantum gates by driving a particular node. The Hamiltonian reads
\begin{eqnarray}
    \label{H_control}
    \bar{\mathcal{H}}&=&\sum_{k=0}^{N}\omega_{k}\ketbra{k}{k}+\Omega(t)\sum_{k>j}(\mathcal{O}_{k,j}\ketbra{k}{j}+\text{H.c}),~~\\
    &&\Omega(t)=\Omega_{X}(t)\cos(\omega_{d}t)+\Omega_{Y}(t)\sin(\omega_{d}t)\nonumber
\end{eqnarray}
where $\omega_{k}$ is the energy of the $k$th energy level, $\mathcal{O}_{k,j}=n_{k,j}+im_{k,j}$ is the matrix element of the charge operator, and $\omega_{d}$ is the driving frequency. In light of the spectrum matrix in Fig.~\ref{fig:schemspect}(a), we observe that the transition $\ket{1}\leftrightarrow\ket{3}$ is the nearest in frequency to the $\ket{0}\leftrightarrow\ket{1}$. Thus, following the typical quantum control procedure, where the system is expressed in the rotating frame, will not capture the physics of our device. Instead, we express the Hamiltonian in the interaction picture with respect to $H=\sum \omega_k-\delta_{k}$, where the latter refers to the detuning between the driving and the transition to be addressed
\begin{eqnarray}
    \label{H_control_RWA}\nonumber
    \mathcal{H}(t)&=&\sum_{k}\delta_{k}\\\nonumber
    &+&\frac{[n_{0,1}+im_{0,1}][\Omega_{X}(t)-i\Omega_{Y}(t)]}{2}\ketbra{0}{1}\\\nonumber
    &+&\frac{[n_{1,3}+im_{1,3}][\Omega_{X}(t)-i\Omega_{Y}(t)]}{2}\ketbra{1}{3}e^{i\alpha t}\nonumber\\
    &+&\rm{H.c},
\end{eqnarray}
where $\alpha=\omega_{3}-2\omega_{1}$ is the energy of the main leakage channel. Here, we focus on the performance of X rotation considering Hahn pulses given by $\Omega_{j}(t)=\Omega_{0,j}\sin^2(\pi t/t_g)$. In this scenario, we choose the amplitude such that the Y rotation on the $\ket{0}\leftrightarrow\ket{1}$ transition vanish, leading to
\begin{eqnarray}
    \label{H_control_RWA2}\nonumber
    \mathcal{H}(t)&=&\sum_{k}\delta_{k}+\frac{\lambda_{1}\Omega_{X}(t)}{2}(\ketbra{1}{0}+\ketbra{0}{1})\\
    & &+\frac{[\lambda_{2}+i\lambda_{3}]\Omega_{X}(t)}{2}\ketbra{1}{3}e^{i\alpha t}+\rm{H.c},
\end{eqnarray}
where $\lambda_{1}=(n_{0,1}^2+m_{0,1}^2)/n_{0,1}$, $\lambda_{2}=\Re{(m_{0,1}+in_{0,1})(m_{1,3}-in_{1,3})}$ and $\lambda_{3}=\Im{(m_{0,1}+in_{0,1})(m_{1,3}-in_{1,3})}$ stand for the dressed matrix elements of the control Hamiltonian. To quantify the performance of our multi-mode device for performing either X or Y gates, we compare it with Transmon and Fluxonium qubits. The comparison is quantified through the gate error $\mathcal{E}(U,V)=1-\mathcal{F}(U,V)$, where $\mathcal{F}$ is the gate fidelity defined as~\cite{Pedersen2007}
\begin{eqnarray}
    \mathcal{F}(U,V) = \frac{\Tr[U_{q}U_{q}^{\dag}]}{d(d+1)} + \frac{|\Tr[U_{q}V^{\dag}]|^2}{d(d+1)}.
\end{eqnarray}
Here, $U_{q}$ represents the truncated unitary operator in the computational subspace, and $V\equiv\{X,Y\}$ is the gate to be implemented. Another important metric corresponds to leakage out of the computational space, defined as 
\begin{eqnarray}
    \mathcal{L}_{\ket{k}}=\frac{1}{d}\sum_{k\neq j =2}^{N}\bigg[|\bra{j}U\ket{k}|^2 + |\bra{k}U\ket{j}|^2\bigg],
\end{eqnarray}
that quantifies the error produced by the outlier transitions. To reduced leakage on the quantum gate we use the pulse shaping techniques of Derivative Removal by Adiabatic Gates (DRAG)~\cite{Motzoi2009}. In this framework, we aim to eliminate leakage contributions through perturbative diagonalization provided by the generator
\begin{eqnarray}
    \label{SWTDRAG}\nonumber
    \hat{S}(t)=\frac{\epsilon\Omega (t)}{2\alpha}\bigg[\beta \lambda_{1}\ketbra{0}{1}+(\lambda_{2}+i\lambda_{3})\ketbra{1}{3}e^{i\alpha t}\bigg]
    -\rm{H.c.},
\end{eqnarray}
where $\epsilon=1/(t_{g}\alpha)$ is the perturbation parameter useful for order counting in the effective Hamiltonian, $\Omega(t)$ is the corrected pulse amplitude, and $\beta$ is a free parameter that controls which type of error DRAG can correct~\cite{Chen2016}. With this generator, the effective second-order Hamiltonian reads
\begin{eqnarray}
    \label{H_eff}
    \mathcal{H}_{\rm{eff}}(t)&=&\mathcal{H}(t)+[\hat{S}(t),\mathcal{H}(t)]\\
    & &+\frac{1}{2}[[\hat{S}(t),\mathcal{H}(t)],\mathcal{H}(t)]+i\dot{\hat{S}}(t),\nonumber
\end{eqnarray}
which at second order in $\epsilon$ has the following structure
\begin{widetext}
    \begin{eqnarray}
        \mathcal{H}_{\rm{eff}}(t)=\begin{bmatrix}
           -\frac{\epsilon^2\beta\lambda_1^2\Omega\Re{\Omega_{X}}}{2\alpha} & \frac{\epsilon\lambda_{1}}{2}\bigg[\Omega_X-i\frac{\beta \dot{\Omega}}{\alpha}\bigg] & 0 & \frac{\epsilon^2(1-\beta)\lambda_1(\lambda_2+i\lambda_3)\Omega\Omega_{X}e^{-i\alpha t}}{4\alpha}\\
           \frac{\epsilon\lambda_{1}}{2}\bigg[\Omega_X^{*}+i\frac{\beta \dot{\Omega}^{*}}{\alpha}\bigg] & \delta_{1}+\frac{\epsilon^2(\beta\lambda_1^2-\lambda_2^2-\lambda_3^2)\Omega\Re{\Omega_{X}}}{4\alpha} & 0 & \frac{\epsilon(\lambda_2+i\lambda_3)}{2}\bigg[\Omega_{X}-\Omega-i\frac{\dot{\Omega}}{\alpha}\bigg]e^{-i\alpha t}\\
           0 & 0 & \delta_2 & 0\\
           \frac{\epsilon^2(1-\beta)\lambda_1(\lambda_2-i\lambda_3)\Omega^{*}\Omega^{*}_{X}e^{i\alpha t}}{4\alpha} & \frac{\epsilon(\lambda_2-i\lambda_3)}{2}\bigg[\Omega^{*}_{X}-\Omega^{*}+i\frac{\dot{\Omega}^{*}}{\alpha}\bigg]e^{i\alpha t} & 0 & \delta_{3}+ \frac{\epsilon^2(\lambda_2^2+\lambda_3^2)\Omega\Re{\Omega_{X}}}{4\alpha}
        \end{bmatrix}~~~~~~
    \end{eqnarray}
\end{widetext}
where $*$ refers to complex conjugate. From the effective Hamiltonian, we can eliminate leakage to the $\ket{1}\leftrightarrow\ket{3}$ transition by choosing the control to be
\begin{eqnarray}
\Omega _{X}(t)=\Omega(t)+i\frac{\dot{\Omega}(t)}{\alpha}.
\end{eqnarray}
\begin{figure}[b]
\includegraphics[width=0.9\linewidth]{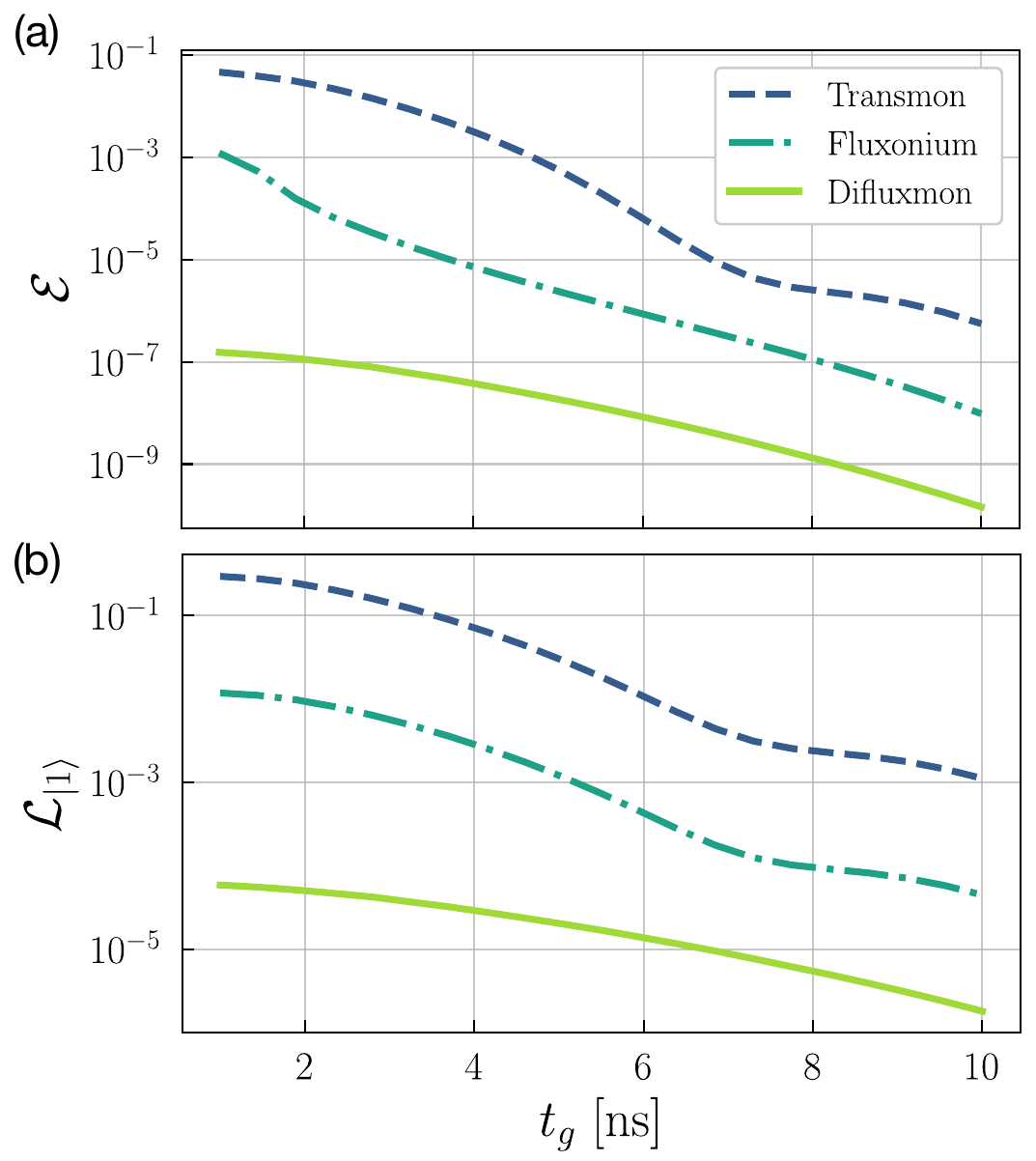}
\caption{\label{fig:dynamics}(a) Gate error $\mathcal{E}$ and (b) leakage $\mathcal{L}_{\ket{1}}$ as a function of the gate time $t_g$ by optimizing the pulse parameters $\{ \Omega_{0}, \delta, \beta \}$ corresponding to the drive amplitude, detuning and DRAG coefficient, respectively.}
\end{figure}
We should note that independently of the value of the parameter $\beta$, the DRAG correction eliminates up to second order in $\epsilon$ for the $\ket{1}\leftrightarrow\ket{3}$ leakage channel. Then, the beta parameter can be set such that we minimize the phase error on the gate. To do so, consider the difference between the diagonal elements of the effective Hamiltonian in the qubit subspace
\begin{eqnarray}
    &\mathcal{H}_{\rm{eff}}[1,1]-\mathcal{H}_{\rm{eff}}[0,0]=
    \delta_{1}+\frac{\epsilon^2(2\beta\lambda_1^2-\lambda_2^2-\lambda_3^2)|\Omega|^2}{4\alpha},
\end{eqnarray}
We can eliminate the phase accumulation by setting $\delta_{k}=k(\delta\epsilon^2)$, obtaining 
\begin{eqnarray}
\delta = -\frac{|\Omega(t)|^2(2\beta\lambda_{1}^2-\lambda_{2}^2-\lambda_{3}^2)}{2\alpha}.
\end{eqnarray}
The validity of $\delta_k$ relies on the effective Hamiltonian, where diagonal terms come from even commutators which scale as $\epsilon^{2j}$. Thus, a time-dependent detuning is able to eliminate (also at second order in $\epsilon$) the phase error on our gate. Another alternative without relying on time-dependent detuning consist in setting $\beta$ such that the phase from the qubit subspace destructively interfere. For our control, the optimal value is given by
\begin{eqnarray}
\label{opt_beta}
\beta = \frac{\lambda_2^2+\lambda_3^2}{2\lambda_{1}^2}.
\end{eqnarray}
For {the} Transmon circuit we have that the control Hamiltonian satisfies $\lambda_1=1$, $\lambda_2=\sqrt{2}$, and $\lambda_3=0$ so that the optimal $\beta$ reducing the phase error is $\beta=1/2$, such variation of the DRAG pulse is known as half-DRAG~\cite{Chen2016}. Notice that either using time-dependent detuning or setting the $\beta$ in the optimal configuration, we should be able to reduce phase error. However, from the effective Hamiltonian, choosing $\beta\neq 1$ induces leakage between the $\ket{0}\leftrightarrow\ket{3}$ transition, which for shorter gating time should also reduce the performance of the gate. For achieving smaller gate error in single-qubit gates below $t_g=10$~ns, we can use typical control strategies such as: considering a non-resonant drive $\omega_d=\omega_{10}+\delta_1$ and find the optimal detuning that cancel out most of the phase error~\cite{Motzoi2013}. Additionally, we can optimize over the drive amplitude $\Omega_{0}$ and DRAG parameter $\beta$ to cancel higher order effects. The numerical optimization were performed using the \texttt{Nelder-Mead} subroutine included in the \texttt{scipy.minimize} package~\cite{NelderMead}. We compare our findings with the same gate optimized for a Transmon circuit with typical parameters $\omega_{10}=2\pi\times 5~$GHz, and $\alpha=-2\pi \times 250~$MHz~\cite{PhysRevA76042319}, and a Fluxonium with parameters $E_{C}=E_{L}= 1.0~$GHz and $E_{J}=4.0~$GHz, leading to $\omega_{10}=2\pi\times 0.58~$GHz, and $\alpha=2\pi \times 3.39~$GHz~\cite{PRXQuantum3037001}. In order to make a fair comparison among all the devices and match experimental capabilities, we bound the parameters $\Omega=2\times \pi(-500, 500)~$MHz, $\delta=2\times \pi(-500, 500)~$MHz and $\beta=(-3,3)$ respectively.
Fig.~\ref{fig:dynamics}(a) and  Fig.~\ref{fig:dynamics}(b) show the gate error $\mathcal{E}$ and the leakage $\mathcal{L}_{\ket{1}}$ as a function of the gate time resulting from the optimization process for all devices. From the figure, we observe a decrease of more than two orders of magnitude in the gate error for our architecture in comparison with Transmon and Fluxonium devices. This improvement arises from the effective cancellation of the principal leakage channel $\ket{1}\leftrightarrow\ket{2}$ in contrast to the Transmon and Fluxonium devices, whose control operators follow a ladder-type structure. In direct comparison with Transmon, our device achieves superior performance owing to a larger anharmonicity associated with the most relevant leakage channel, featuring an anharmonicity three times larger than that observed in Transmon. Regarding Fluxonium, although its anharmonicity is larger, its matrix elements and frequency are approximately three times smaller than those featured by our device. This difference enables high-fidelity gate operations even with bounded drive {strengths}, suppressing counter-rotating errors that arise at low qubit frequencies. Our optimized dynamics yield a gate error of approximately $10^{-7}$ for an X gate at a gate time of $5$~ns. Furthermore, while we observe a reduction in leakage, this decrease is less pronounced than the improvement in gate error. Consequently, we conclude that for short gate times leakage may limit the qubit performance.
{It is worth noting that the results presented here correspond to the unitary limit of the gate simulations and do not include incoherent errors arising from finite coherence times or other noise sources. The inclusion of such effects is expected to increase the total error. However, given the short gate durations and moderate drive strengths considered, these incoherent contributions are expected to remain subdominant.}
\section{Readout and active reset} \label{sec:Readoutreset}
\begin{figure*}
\includegraphics[width=\linewidth]{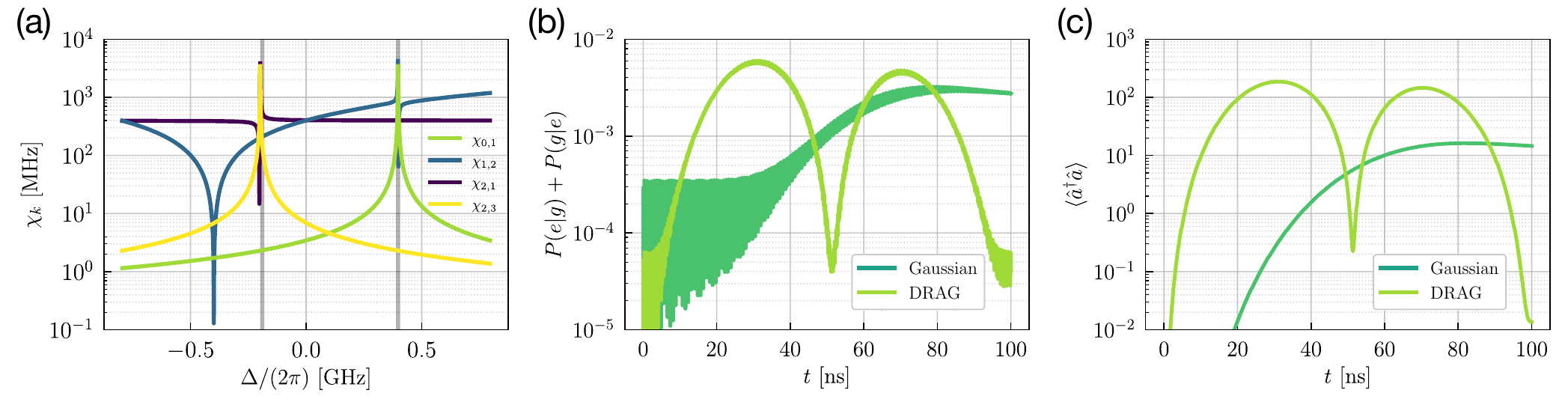}
\caption{\label{fig:Reset}(a) Dispersive Stark shift $\bar{\chi}_{k,\ell}$ of the lowest two Difluxmon energy levels as a function of the resonator-drive detuning $\Delta$. Vertical lines corresponds to the first two energy transition of the artificial atom, where the shift diverges. (b) evolution of the assignment error of the Difluxmon states for a readout using gaussian pulse and using DRAG corrected pulse. (c) Photon number stored on the resonator as a function of the measurement time for the Gaussian and the DRAG corrected pulses. The simulation parameters are given by $\omega_r=2\pi\times 6.99~$GHz, $\epsilon_{10}=2\pi\times2.5~$GHz, {$g=2\pi\times 37$~MHz and $\kappa = 2\pi \times 1.32~\text{MHz}$.}}
\end{figure*}
The next step in the characterization of our multi-mode qubit corresponds to the study of the dispersive readout when it is coupled to a transmission line resonator~\cite{PhysRevA69062320}. We denote as $\ket{k}$ and $\epsilon_{k}$ to the $k$th states and energies of the circuit, respectively. In this basis, the Hamiltonian reads
\begin{eqnarray}
\label{Hamiltonian_transmon_diag}
&\mathcal{H}=&\sum_{k}\epsilon_{k}\ketbra{k}{k}+\omega_{\rm{r}}\hat{a}^{\dag}\hat{a}+\Omega(t)\cos(\omega_{r}t)(\hat{a}^{\dag}+\hat{a})\nonumber\\
& & +\sum_{k,k'}g_{k,k'}\ketbra{k}{k'}(\hat{a}^{\dag}+\hat{a}),
\end{eqnarray}
here, $g_{k,k'}=g\bra{k}\hat{n}_{3}\ket{k'}/|\bra{0}\hat{n}_3\ket{1}|$ is the dressed charge coupling strength for the node $\{3\}$ and $\Omega(t)$ the envelope of the microwave drive used for the readout. We can simplify the Hamiltonian by expressing it in the rotating frame with respect to the drive frequency, and eliminate the counter-rotating terms through the rotating wave approximation, leading to the generalized Jaynes-Cummings model
\begin{eqnarray}
\label{GJCM}\nonumber
\mathcal{H}_{{\rm{JCM}}}&=&\sum_{k}(\epsilon_{k}-\omega_{\rm{r}}k)\ketbra{k}{k}+\frac{\Omega(t)}{2}(\hat{a}^{\dag}+\hat{a})\\
 & &+\sum_{k}g_{k}(\ketbra{k}{k+1}\hat{a}^{\dag}+\ketbra{k+1}{k}\hat{a}).
\end{eqnarray}
Considering common resonator characteristics~\cite{Pfeiffer2023}, for a resonator frequency of $\omega_{\rm{r}}=2\pi\times 6.990$~GHz, we obtain cavity-qubit detuning $\Delta_{10}=|\omega_{1}-\omega_{0}-\omega_{\rm{r}}|=2\pi\times4.502$~GHz which, for coupling strength $g_{1}=2\pi\times37$~MHz {and $\kappa = 2\pi \times 1.32~\text{MHz}$}, allows us to operate in the dispersive regime, where the Hamiltonian is diagonal up to second order in the expansion parameter $\bar{\chi}_1=g^2/\Delta_{10}$
\begin{eqnarray}
\label{HEFF}\nonumber
\mathcal{H}_{{\rm{eff}}}&=&\sum_{k}(\epsilon_{k}-\omega_{\rm{r}}k)\ketbra{k}{k}+\sum_{k}\chi_{k}\ketbra{k}{k}\hat{a}^{\dag}\hat{a}\\
 & &+\frac{\Omega(t)}{2}(\hat{a}^{\dag}+\hat{a}).
\end{eqnarray}
Here, $\chi_{k}=\sum_{\ell}[\bar{\chi}_{k,\ell}-\bar{\chi}_{\ell,k}]$ is the total dispersive shift taking into account the multi-level nature of our device, with $\bar{\chi}_{\ell,k}=|g_{\ell,k}|^2/(\epsilon_{\ell,k}-\omega_{\rm{r}})$ being the state-dependent stark-shift. Notice that for artificial atoms truncated in the qubit subspace, we obtain the typical Stark-shift of the form $\chi_{1}=\bar{\chi}_{1}$. Nevertheless, {accounting for higher-lying energy levels modifies the Stark shift to} $\chi_{1}=g_{1}^2/\Delta_{10}-g_{2}^2/(2\Delta_{21})$~\cite{PhysRevLett105223601}, obtaining a smaller contrast between the computational states. Fig.~\ref{fig:Reset}(a) shows the relevant Stark shift that contributes to $\chi_{0}$ and $\chi_{1}$ to measure the states $\ket{0}$ and $\ket{1}$, respectively. Thus, we need to choose the {appropriate} node for implementing the readout. Thanks to the multi-mode nature of the device we can choose different nodes for different operations. In particular in our device, we set the node \{1\} for driving purposes while the remaining could be used for readout by coupling either an additional resonator with Purcell filters at the qubit frequency~\cite{Sete2015}. In light of Fig.~\ref{fig:matelem}(a) we see that the matrix elements of the node \{3\} have larger values than the second one, so that the capacitive coupling is stronger allowing to achieve better contrast $\chi$ and consequently faster readout.
To understand how the measurement works, let us consider the Heisenberg equation of motion for the cavity operator $\hat{a}$. In this scenario, the equation of motion reads
\begin{eqnarray}
\label{input_output_eq}
\langle\dot{\hat{a}}\rangle&=&i\chi_k\langle\hat{a}\rangle-\frac{\kappa}{2}\langle\hat{a}\rangle+i\Omega(t),
\end{eqnarray}
where $\kappa$ is the decay rate of the resonator. For a constant drive, $\Omega(t)=\Omega$, and the resonator initialized in the vacuum, we obtain an steady state solution of the form
\begin{eqnarray}
\label{input_output_sol}
\langle\hat{a}(t)\rangle=-\frac{4\chi_{k}\Omega}{\kappa^2+4\chi_{k}^2}+i\frac{2\kappa\Omega}{\kappa^2+4\chi_{k}^2},
\end{eqnarray}
{which} represents the position of the multi-level system in the $\rm{IQ}$ plane of the resonator. Thus, different states may arrive to different points of such plane, which could be maximized depending on the microwave envelope to be used. In our case, we consider a pulse of the form $\Omega(t)=\Omega_{0}\sin^3(\pi t/t_{m})$, where $\Omega_{0}$ is the pulse strength and $t_{m}=100$~ns is the measurement time. We calibrate $\Omega_{0}=2\pi\times35.39~$MHz such that the number of {photons} stored in the resonator {is} equal to five at the {midpoint} of the dynamics. We characterize the readout process according to two key criteria: (i) a negligible photon number stored in the resonator after the readout process, ensuring that the cavity is effectively empty for reuse, and (ii) minimal assignment error, such that a qubit initialized in state $\ket{g}$ (or $\ket{e}$) remains in that state after the readout. Fig.~\ref{fig:Reset}(b) displays the photon number in the resonator for initializations in $\ket{g}$ and $\ket{e}$. From the figure, we observe two key aspects. First, the resonator is driven in such a way that its critical photon number {$\bar{n}_{\rm{crit}}\equiv[\Delta_{10}/(2g_{1})]^2\equiv11\times10^3$} is not exceeded. This ensures that the multi-mode system is not driven outside the potential, avoiding ionization phenomena~\cite{Ionization}, and mitigating the influence of nonlinearities in the measurement outcome~\cite{PhysRevA77060305}. Secondly, we observe that, although the envelope vanishes at $t=t_{m}$, the {residual} photon number does not decrease during the readout {due to the slow decay rate of the resonator}. Consequently, for an {effective} use of the multi-mode qubit {and readout resonator in the described configuration}, it is necessary to implement an active reset mechanism. 
The reset mechanism to be analyzed, DRACHMA, is the same as introduced in Ref.~\cite{Motzoi2018} and experimentally demonstrated Ref.~\cite{jerger2024dispersive}, which relies on using DRAG-like pulses to reverse engineer the resonator's transfer function, guaranteeing at the same time reduced leftover photon and qubit population. The starting point is the input-output equation of motion for the resonator operator, expressed in the Fourier domain
\begin{eqnarray*}
\label{input_output_eq_freq}
i\omega\bar{a}(\omega)&=&i\chi_k\bar{a}(\omega)-\frac{\kappa}{2}\bar{a}(\omega)+i\bar{\Omega}(\omega)-\sqrt{\kappa}\bar{a}_{{\rm{in}}}(\omega),
\end{eqnarray*}
where the bar variables are the Fourier transformed function, and $\bar{a}_{{\rm{in}}}(\omega)$ is the input field on the transmission line resonator. We have many input-output relation as states in our multi-mode qubit. In our derivation, we will fix a single value of $\chi_{k}$, and the generalization to higher states is straightforward. In absence of the drive, we write the input field as
\begin{eqnarray}
\bar{a}_{{\rm{in}}}(\omega) = \frac{1}{\sqrt{\kappa}}\bigg[i(\omega-\chi_k)+\frac{\kappa}{2}\bigg]\bar{a}(\omega),
\end{eqnarray}
That corresponds to the inverse of the transfer-function of the qubit-resonator transfer function $T(\omega)_{\ket{k}}^{-1}$. If we choose the drive envelope to be 
\begin{eqnarray}
\bar{\Omega}(\omega) = -\frac{i}{\sqrt{\kappa}}\bigg[i(\omega-\chi_k)+\frac{\kappa}{2}\bigg]\bar{\Omega}_{{\rm{trial}}}(\omega),
\end{eqnarray}
where $\bar{\Omega}_{{\rm{trial}}}(t)=\Omega_{0}\sin^3(\pi t/t_{m})$ is a trial function to be optimized. This selection allows us to obtain zero output field $\bar{a}(\omega)=0$ at the end of the measurement time when the system is initialized in the state $\ket{k}$. For correcting several qubit states, the pulse envelope reads
\begin{eqnarray}
\bar{\Omega}(\omega) = \prod_{k=0}^{N}T(\omega)_{\ket{k}}^{-1}\bar{\Omega}_{{\rm{trial}}}(\omega),
\end{eqnarray}
Fig.~\ref{fig:Reset}(b,c) shows the evolution of the photon number and the error assignment {$[P(e|g)+P(g|e)]$}. We observe a photon number of {$\langle \hat{a}^\dagger\hat{a}\rangle\approx 10^{-2}$} at the end of the readout process demonstrating an empty cavity, which differs with the {$\langle \hat{a}^\dagger\hat{a}\rangle\approx 10^{1}$} present in absence of control. Furthermore, we see a decrease of two orders of magnitude in the error assignment. More precisely, using DRACHMA pulses we expect errors of $1.39\times10^{-4}$ and $4.12\times10^{-4}$ for the readout and reset, whereas without pulse shaping we obtain errors of $9.07\times10^{-4}$ and $27.6\times10^{-4}$ for the readout and reset, respectively. Additionally, we can also quantify the induced dephasing rate due to shot noise given by $\Gamma_{2,m}=-2\chi_{0}\Im[\alpha_{\ket{0}}\alpha_{\ket{1}}^{*}]$~\cite{PhysRevA.74.042318} where $\alpha_{\ket{\ell}}$ is the field amplitude when the qubit is initialized in the state $\ket{\ell}$. We observe that with the DRAG pulse $\Gamma_{2,m}=2\pi \times 0.02233$ kHz while without pulse shaping the dephasing rate is $\Gamma_{2,m}=2\pi \times 27.14$ kHz which is again almost two order of magnitude larger than the one obtained with pulse shaping. {Another critical aspect that has not been analyzed so far is the $T_1$ limit set by Purcell decay into the readout resonator. For the set of parameters considered here, the qubit–resonator system can be assumed to operate well within the dispersive regime. In this case, the Purcell-induced decay rate can be estimated as $\gamma = \left(\frac{g}{\Delta}\right)^2\kappa$~\cite{Houck2008}. This corresponds to a decay time on the order of milliseconds, suggesting that Purcell decay does not constitute a dominant limitation on the $T_1$ time in the present parameter regime.}
\section{Resilience to parameter fluctuations} \label{sec:resilience}
\begin{figure}[b]
\includegraphics[width=0.9\linewidth]{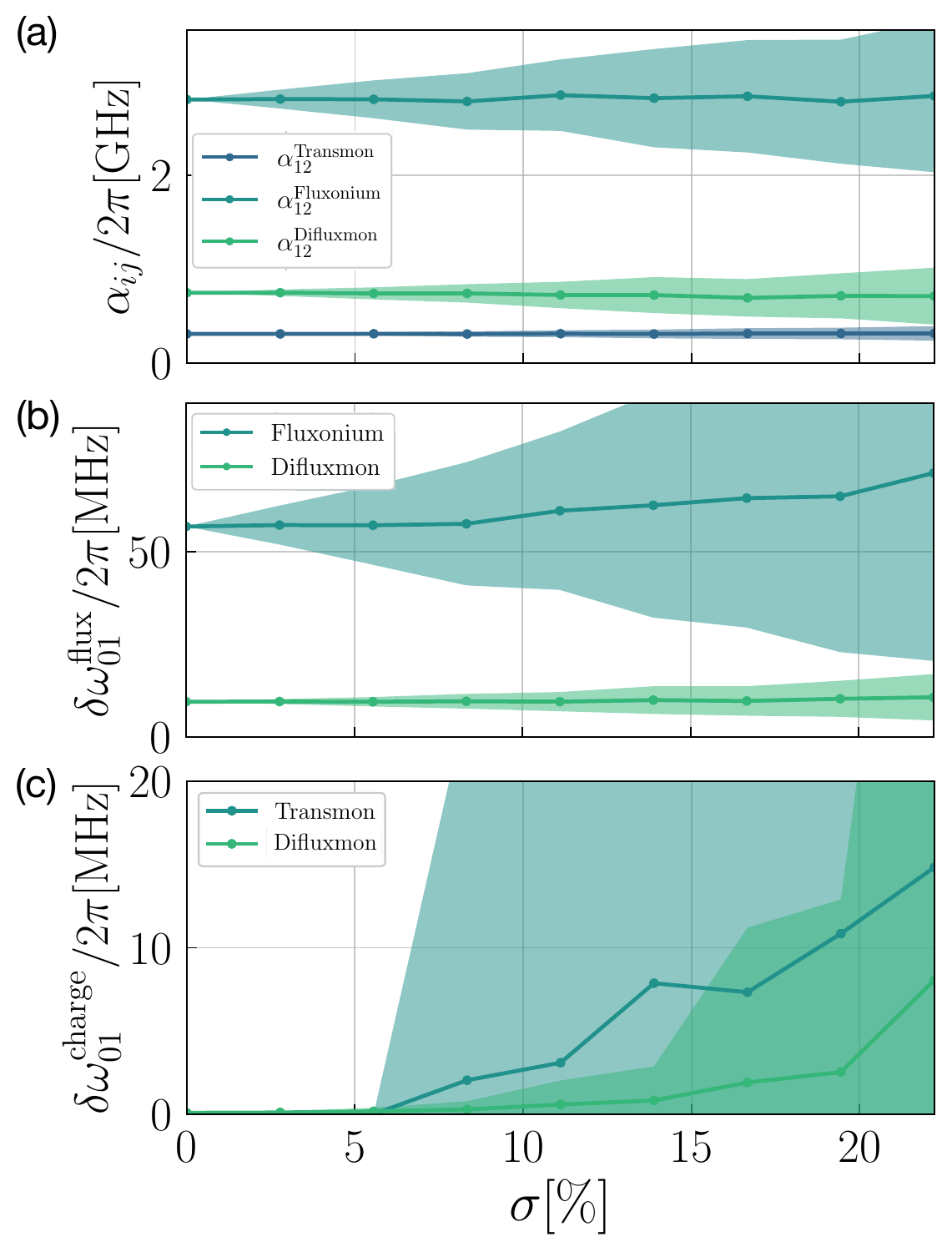}
\caption{\label{fig:resilience} Resilience to parameter fluctuations simulation. Mean values and standard deviation of (a) anharmonicity $\alpha_{ij}$, (b) energy dispersion for $10^{-2} \Phi_0$ flux fluctuations $\delta \omega_{01}^{\rm flux}$, and (c) energy dispersion for $e^{-}$ charge fluctuations $\delta \omega_{01}^{\rm charge}$; for 500 circuits instances with values taken from a normal distribution around the optimal physical values considering different standard deviation $\sigma$ of the normal distributions.}
\end{figure}
{The proposed device requires a specific parameter regime in order to exhibit the desirable characteristics discussed in the previous sections.} For that reason, in this section we focus on studying how fluctuations in physical values of the components constituting the circuit, expected from fabrication inaccuracies, modify the characteristics of the device. For that purpose, we computed the system characteristics for multiple circuit instances modifying the capacitors, Josephson junction energies, and linear inductors, considering normal distributions around the optimal values for different degrees of standard deviation. In Fig.~\ref{fig:resilience} we show the variation of the anharmonicity, charge and flux dispersion under different degrees of deviation, compared to Transmon and Fluxonium devices. We observe that under reasonable values of deviation expected from fabrication, the desired characteristics of the device are preserved, thus accounting for certain fabrication resilience of the obtained solution. As a proof of concept on fabrication, in Appendix~\ref{sec:appendixC} section we present a lithographic proposal of the device, optimized to fulfill the capacitive relations obtained in the theoretical design optimization, where the capacitive relations are matched with an estimated error smaller than $\sim 10 \%$. This capacitive mismatch could also be decreased by considering the extra capacitance induced by the introduction of the junction and junction arrays to the design, so that a lower-shooting approach to match the capacitive relations could be beneficial.
\section{Towards Scalability} \label{sec:scalability}
A crucial aspect to consider when engineering superconducting qubits is scalability. Proposals for novel superconducting qubits have primarily focused on improving the coherence time of the computational space by reducing the charge (flux) matrix elements involving the computational states. This approach reduces the coupling of the system to both the environment and other circuit components. Although this method effectively improves coherence, it has the side effect of significantly reducing controllability. Although this reduction in matrix elements can be compensated by increasing the driving {strength}, it may introduce significant issues such as heating~\cite{Cohen2023,Ateshian2025} or ionization~\cite{Dumas2024}. Therefore, it is important to balance the coupling elements so that the device remains protected yet addressable. This effect is even more significant when considering two-qubit operations, as the coupling between the qubits decreases quadratically with the reduction in matrix elements. This is a crucial limiting factor in scaling protected devices such as (Heavy) Fluxonium or Zero-$\pi$, which ultimately rely on the application of CPhase operations that also lead to parasitic interactions \cite{Brooks2013}. Conversely, for the Transmon, the large matrix elements facilitate its coupling to other qubits; however, its relatively limited \( T_1 \) times constrain the total coherence times.

{As an illustrative example, we consider the cross-resonance gate in fixed-frequency architectures. In Transmon qubits, large charge matrix elements yield strong XZ interactions but also sizable static ZZ terms that produce crosstalk. The Difluxmon qubit instead exhibits charge matrix elements approximately a factor of two smaller than those of a Transmon, while remaining about three times larger than in experimentally demonstrated Fluxonium implementations~\cite{Dogan2023}. This intermediate regime is expected to mitigate static ZZ interactions while still allowing fast two-qubit gates, at the cost of moderately increased drive amplitudes.}

The balance found for the matrix elements of the qubit proposed here takes a step forward toward finding proper configurations suitable for the implementation of two-qubit operations, while maintaining sufficiently relevant anharmonicity and coherence times, thereby decreasing the ratio between the gate time and the total coherence time.
\section{Conclusion}
In this article, we have presented a qubit {based on} a multi-mode superconducting circuit {operated in a specific parameter regime, which we refer to as the Difluxmon qubit}. We have compared its characteristics with two of the most extensively studied superconducting qubits, Transmon and Fluxonium. We have demonstrated that the characteristics of the Difluxmon qubit achieve a balance between Transmon's controllability and resilience to dephasing effects, and Fluxonium's anharmonicity and resistance to depolarization. To quantify this performance, we consider usual coherence times and microwave charge control gate times for $\pi$ rotations with error $\mathcal{E}<10^{-6}$. Considering Transmon coherence and gate times  of ($T_2 \sim 200\,\mu\text{s}$, $t_g \sim 20\,\text{ns}$)~\cite{Hyyppa2024}, we compute a fundamental limit of operations to be $T_2/t_g \sim 10^4$. Similarly for Fluxonium systems, reasonable coherence and gate times ($T_2 \sim 800\,\mu\text{s}$, $t_g\sim 10\,\text{ns}$)~\cite{Rower2024} would lead to $T_2/t_g \sim 8 \cdot 10^4$, offering an eightfold improvement over Transmon. For the device presented here, we estimate coherence and gate times ($T_2\sim400\,\mu\text{s}$, $t_g \sim 2\,\text{ns}$) leading to $T_2/t_g \sim 20\cdot10^4$, offering more than a twofold improvement over Fluxonium device. This comparison does not account for additional critical factors, such as coupling to neighboring qubits, readout performance or other gate schemes, such as flux controlled gates. However, based on the computed matrix elements, we expect the Difluxmon qubit to perform at least equally than other depolarization-resistant qubits. Future studies should focus on evaluating the experimental feasibility of the device—some prospects for which are discussed in Appendix~\ref{sec:appendixC}—, including other possible decoherence channels that could limit the coherence time, as well as its scalability and the performance of two-qubit operations.
\begin{acknowledgments}
The authors would like to thank Frederik Pfeiffer and Christian Schneider for useful discussions and insightful comments. The authors acknowledge support from OpenSuperQ+100 (Grant No. 101113946) of the EU Flagship on Quantum Technologies. This project has also received support from the Spanish Ministry of Economic Affairs and Digital Transformation through the QUANTUM ENIA project call - Quantum Spain, and by the EU through the Recovery, Transformation and Resilience Plan–NextGeneration EU within the framework of the Digital Spain 2026 Agenda, and Basque Government through Grant No. IT1470-22, and through the Elkartek project KUBIBIT-kuantikaren berrikuntzarako ibilbide teknologikoak (ELKARTEK25/79). P. G. A. acknowledges support from UPV/EHU Ph.D. Grant No. PIFG 22/25. F. A. C. L acknowledges support from the German Ministry for Education and Research, under QSolid, Grant no. 13N16149. M. S. acknowledges support from Project Grant No. PID2024-156808NB-I00 and Spanish Ramón y Cajal Grant No. RYC-2020-030503-I funded by MICIU/AEI/10.13039/501100011033 and by “ERDF A way of making Europe” and “ERDF Invest in your Future”, and from the IKUR Strategy under the collaboration agreement between Ikerbasque Foundation and BCAM on behalf of the Department of Education of the Basque Government. G. R. acknowledges Dicyt USACH under grant 5392304RH-ACDicyt and Financiamiento Basal para Centros Cient\'ificos y Tecnol\'ogicos de Excelencia (Grant No. AFB220001).

\end{acknowledgments}
\appendix
\section{Hamiltonian Derivation}\label{sec:appendixA}
The Lagrangian description of the circuit presented in Fig.~\ref{fig:schemspect}(a), considering node $\{0\}$ as reference, is given by
\begin{eqnarray}\nonumber
    \hat{\mathcal{L}} &=&\, \frac{1}{2} \dot{\boldsymbol{\phi}}^T \boldsymbol{C} \dot{\boldsymbol{\phi}} - \frac{1}{2} \boldsymbol{\phi}^T \boldsymbol{L}^{-1} \boldsymbol{\phi}\\
    & &+ E_J^{b_0} \cos{\left(\frac{2\pi}{\Phi_0} \hat{\phi}_1\right)}\nonumber\\
    & &+ E_J^{b_5} \cos{\left(\frac{2\pi}{\Phi_0}\left( \hat{\phi}_2 - \hat{\phi}_3\right)\right)},
\end{eqnarray}
where $\boldsymbol{\phi}^T = (\hat{\phi}_1, \hat{\phi}_2,\hat{\phi}_3)$ is the flux-node vector. The capacitance and inductance matrices are given by 
\begin{eqnarray*}
    \boldsymbol{C} =
    \scalebox{0.9}{$
    \begin{pmatrix}
        C^{b_0}+C^{b_3}+C^{b_4} & -C^{b_3} & -C^{b_4}\\
        -C^{b_3} & C^{b_1}+C^{b_3}+C^{b_5} & -C^{b_5}\\
        -C^{b_4} & -C^{b_5} & C^{b_2}+C^{b_4}+C^{b_5}\\
    \end{pmatrix}
    $}
\end{eqnarray*}
and
\begin{eqnarray*}
    \boldsymbol{L}^{-1} = \begin{pmatrix}
        1/L_{b^4} & 0 & -1/L_{b^4}\\
        0 & 0 & 0\\
        -1/L_{b^4} & 0 & 1/L_{b^2}\!\!+1/L_{b^4}
    \end{pmatrix}.
\end{eqnarray*}\\
\normalsize
Performing a Legendre transform we can obtain the Hamiltonian of the system, given by
\begin{eqnarray}
    \hat{H} &=& \, \frac{1}{2} \boldsymbol{q}^T \boldsymbol{C}^{-1} \boldsymbol{q} + \frac{1}{2} \boldsymbol{\phi}^T \boldsymbol{L}^{-1} \boldsymbol{\phi}\nonumber\\
    & &- E_J^{b_0} \cos{\left(\frac{2\pi}{\Phi_0} \hat{\phi}_1\right)}\nonumber\\
    & &- E_J^{b_5} \cos{\left(\frac{2\pi}{\Phi_0}\left( \hat{\phi}_2 - \hat{\phi}_3\right)\right)}.
\end{eqnarray}
By performing the change to dimensionless variables $\hat{q} = -2e\, \hat{n}$ and $\hat{\phi} = \frac{\Phi_0}{2\pi}\, \hat{\varphi}$, we can rewrite
\begin{eqnarray}\label{eq:HamApendix}
    \hat{H} &=&\,\, 4\, \boldsymbol{n}^T \boldsymbol{E}_C\, \boldsymbol{n} + \frac{1}{2}\, \boldsymbol{\varphi}^T \boldsymbol{E}_L\, \boldsymbol{\varphi}\nonumber\\
    & &- E_J^{b_0} \cos{\left( \hat{\varphi}_1\right)}\nonumber\\
    & &- E_J^{b_5} \cos{\left( \hat{\varphi}_2 - \hat{\varphi}_3\right)},
\end{eqnarray}
with ${[E_C]}_{ij} = \frac{e^2}{2}{[C^{-1}]}_{ij}$ and ${[E_L]}_{ij} = \left(\frac{\Phi_0}{2\pi}\right)^2{[L^{-1}]}_{ij}$. In order to complete the description, we should include the effect of external biases on the system. In every node of the system $ N_i$ with $i \in \{1,2,3\}$ we may experiment an external charge bias ${[n_{g_{ext}}]}_i$. Considering this bias to be static in time, we can always make a proper choice of gauge to eliminate it in "flux-like" degrees of freedom \cite{PhysRevLett103217004}, keeping only the contribution to the "charge-like" degree of freedom of our system $N_2$, so that we can represent its effect by making the substitution $\hat{n}_2 \rightarrow \hat{n}_2 + {n_g}_{ext}$ in the Hamiltonian description in Eq.~\ref{eq:HamApendix}. For the introduction of the external flux effect to the model, we should consider every inductive loop in the system. The single inductive loop in the device is the one produced by the series connection of the branches $\{b_0, b_4, b_2\}$. To reflect the effect of external magnetic flux bias threading the loop we include an extra static constant $\varphi_{ext}$ in the linear inductor term of the branch $b_2$, selected as the one closing the loop. All the described effects can be modeled, including the term
\begin{eqnarray}
    \hat{H}_{\text{ext}} &=&\, 8\left( [E_C]_{11}\hat{n}_2 + [E_C]_{01}\hat{n}_1 + [E_C]_{12}\hat{n}_3 \right) n_{g_{\text{ext}}}\nonumber\\
    & & + [E_L]_{b_2}\hat{\varphi}_3\,\varphi_{\text{ext}},
\end{eqnarray}
to the Hamiltonian description in Eq. \ref{eq:HamApendix}.
\section{Numerical simulation} \label{sec:appendixB}
To implement the Hamiltonian numerically, we distinguish between \textit{charge-like} or periodic modes, and \textit{flux-like} or extended modes. The former ones are represented in the so-called charge basis where operators are represented as
\begin{eqnarray}
    \hat{n} &=& \sum_{n=-N_c}^{N_c} n\, |n\rangle \langle n|,\\\nonumber
    \cos{\left(\hat{\varphi}\right)} &=& \sum_{n=-N_c}^{N_c} \frac{1}{2} \left(|n+1\rangle \langle n| + |n\rangle \langle n+1|\right),\\\nonumber
    \sin{\left(\hat{\varphi}\right)} &=& \sum_{n=-N_c}^{N_c} \frac{1}{2i} \left(|n+1\rangle \langle n| - |n\rangle \langle n+1|\right).
\end{eqnarray}
Additionally, for the estimation of the coherence times due to quasiparticle tunneling effects, we need to redefine the charge and the phase operator, where instead of representing number and tunneling of Cooper pairs, now it represents the number and tunneling of electrons $\ket{n^{\text{SE}}}$~\cite{PhysRevB84064517}. In this representation, the relevant operators are expressed as follows
\begin{eqnarray}
    \hat{n} &=&\!\! \sum_{n=-N_{E,c}}^{N_{E,c}}  n^{\text{SE}}\, |n^{\text{SE}}\rangle \langle n^{\text{SE}}|,\\\nonumber
    \cos{\left(\hat{\varphi}\right)} &=& \!\!\sum_{n=-N_{E,c}}^{N_{E,c}} \frac{1}{2} \left(|n^{\text{SE}}+2\rangle \langle n^{\text{SE}}| + |n^{\text{SE}}\rangle \langle n^{\text{SE}}+2|\right),\\\nonumber
    \sin{\left(\hat{\varphi}\right)} &=& \!\!\sum_{n=-N_{E,c}}^{N_{E,c}} \frac{1}{2i} \left(|n^{\text{SE}}+2\rangle \langle n^{\text{SE}}| - |n^{\text{SE}}\rangle \langle n^{\text{SE}}+2|\right),\\\nonumber
    \cos{\left(\frac{\hat{\varphi}}{2}\right)} &=& \!\!\sum_{n=-N_{E,c}}^{N_{E,c}} \frac{1}{2} \left(|n^{\text{SE}}+1\rangle \langle n^{\text{SE}}| + |n^{\text{SE}}\rangle \langle n^{\text{SE}}+1|\right),\\\nonumber
    \sin{\left(\frac{\hat{\varphi}}{2}\right)} &=& \!\!\sum_{n=-N_{E,c}}^{N_{E,c}} \frac{1}{2i} \left(|n^{\text{SE}}+1\rangle \langle n^{\text{SE}}| - |n^{\text{SE}}\rangle \langle n^{\text{SE}}+1|\right).
\end{eqnarray}
On the other hand, to obtain the numerical representation of the flux operator for modes represented in the charge basis, we obtained an expression relating the flux operator of the mode with all the charge operators of the device, computing the commutator \cite{PhysRevX11011010}
\begin{eqnarray}\nonumber
    \bra{i}{[\hat{\varphi}_m, \hat{H}]}\ket{j} =\, \hbar\omega_{ij} \bra{i}\hat{\varphi}_m \ket{j}
    = \sum_k 8i\,[E_C]_{km} \bra{i}\hat{n}_k \ket{j}
\end{eqnarray}
For the implementation of operators representing \textit{flux-like} modes, we made use of the harmonic oscillator basis, considering the linear part of each mode to construct the numerical representation of the operators by means of creation and annihilation operators such that
\begin{equation}
\begin{split}
    \hat{n}_m &= \frac{i}{\sqrt{2}}\left( \frac{[E_L]_{mm}}{8[E_C]_{mm}} \right)^{\frac{1}{4}} \left( \hat{b}_m - \hat{b}_m^\dagger \right)\\
    \hat{\varphi}_m &= \frac{1}{\sqrt{2}} \left(\frac{8[E_C]_{mm}}{[E_L]_{mm}} \right)^{\frac{1}{4}} \left( \hat{b}_m + \hat{b}_m^\dagger \right)
\end{split}
\end{equation}
In order to assure convergence of the numerical results, convergence test were conducted, leading to consider the cutoffs of $N_c = 20$ for charge-like modes and $N_f = 41$ for flux-like modes.
\section{Wave function representation}\label{sec:appendixW}
As described in the numerical methods section, the representation of the system is performed expressing the periodic mode two in charge basis $N \in [-N_{c},N_{c}]$, and extended modes one and three in Fock basis $\{n,m\} \in [0,N_f]$. To go from the original eigenstates obtained from diagonalization $|\Psi\rangle = \sum_{nNm} C_{nNm}\,|n\rangle |N\rangle|m\rangle$ to the flux space representation we use
\begin{equation}
\psi(\varphi_1,\varphi_2,\varphi_3) = \sum_{n,N,m} C_{nNm} \langle\varphi_1|n\rangle\langle\varphi_2|N\rangle\langle\varphi_3|m\rangle,
\end{equation}
where flux space representation of Fock states is given by 
\begin{eqnarray}
\langle\varphi_{i}|k\rangle &=& \frac{1}{\sqrt{2^k k!}}\left(\sqrt{\frac{[E_L]_{ii}}{8[E_C]_{ii}}}\frac{1}{\pi}\right)^{\frac{1}{4}}e^{-\sqrt{\frac{[E_L]_{ii}}{8[E_C]_{ii}}}\frac{\varphi_i^2}{2}}\nonumber\\
& &\times H_k\left(\left( \frac{[E_L]_{ii}}{8[E_C]_{ii}}\right)^{\frac{1}{4}}\varphi_i\right),
\end{eqnarray}
for $i \in \{1,3\}$ and $k \in \{n,m\}$. The representation of the charge states in flux space is given by the discrete Fourier transform
\begin{equation}
    \langle\varphi_2|N\rangle = \frac{1}{\sqrt{2N_c+1}}\sum_Ne^{i\varphi_2\frac{N}{2N_c+1}}.
\end{equation}
Similarly, to obtain the charge space wave functions $\phi_i$ we perform the Fourier transform of the obtained flux space representation. For the projections depicted in Fig.~\ref{fig:schemspect}(b-c) we used $\varphi_{1,3} \in [-5\pi/2,\pi/2]$ and $\varphi_2 \in [0,2\pi]$, and traced out one of the dimensions to get a two-dimensional representation.
\section{Lithographic Implementation} \label{sec:appendixC}
\begin{figure}[b]
\includegraphics[width=\linewidth]{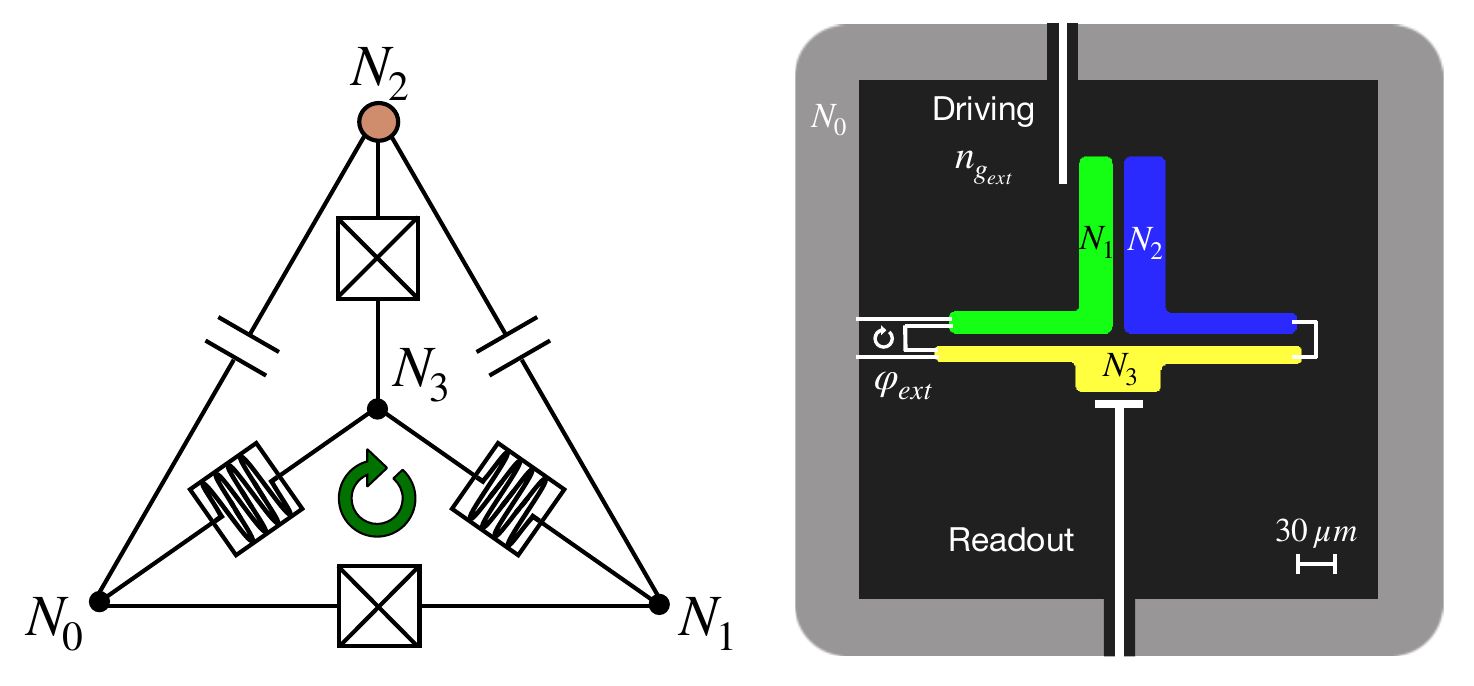}
\caption{\label{fig:lithography} Proposed lithographic implementation for the optimized device. Lumped-element description of the device (top) followed by the geometric design (bottom) with indicated: nodes $N_i$, links between islands, driving and readout points. The colored part represents the optimized geometry, while the white parts act as schematics of branches and driving/readout structures.}
\end{figure}
In the optimization process, special attention was put into evaluating designs in a range of physical parameters possible to implement in an experimental setup. As a proof of concept on the possibilities of fabrication of the device, here we propose a lithographic implementation of the system. The geometry was optimized using evolutionary optimization methods to best suit the capacitive relations between circuit islands, still taking into consideration usual fabrication constraints. The procedure consisted of generating different parameterized geometrical layouts, which were optimized using evolutionary methods, numerically computing the capacitive relations between islands using a fast field solver~\cite{Fastcap1,Fastcap2}, and optimizing the parameters looking to match the capacitive relations of the theoretical design. For the evaluation of the degree of similarity between the theoretical and the lithographic design, the Frobenius norm of the matrix difference $\gamma$ \small $= \sqrt{\sum_{ij} \left(\boldsymbol{C}_{ij}^{\text{theo}} - \boldsymbol{C}_{ij}^{\text{litho}}\right)^2}$ \normalsize was chosen. For simple geometrical designs, such as the one shown in Fig. \ref{fig:lithography}, similarities of around $\gamma \approx 8.5$ fF were achieved, showing that the device expresses realistic capacitive relations for experimental setups.
\section{Evolutionary algorithm} \label{sec:appendixD}
\begin{figure}[b]
\includegraphics[width=\linewidth]{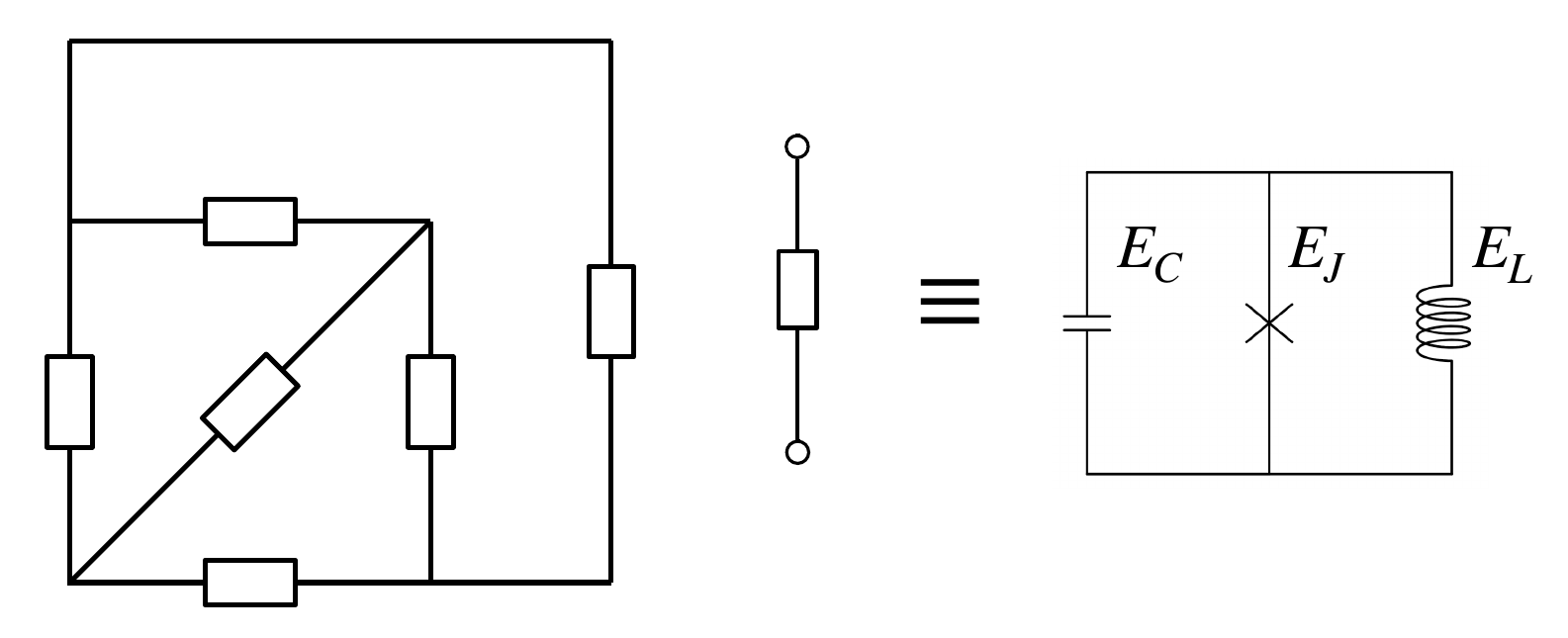}
\caption{\label{fig:optimization}Circuit structure considered of $N=4$ nodes with all-to-all connections where each branch contains a capacitance and, depending on the configuration, a linear and(or) a non-linear inductor.}
\end{figure}
{The task of identifying superconducting architectures exhibiting specific system characteristics can be described as a double optimization over both discrete and continuous parameter spaces.} The algorithm must find suitable topologies by changing the arrangement (discrete) of the circuit component while finding the optimal circuit parameters (continuous) such that the device {gives} rise to the targeted features. This problem is {computationally} demanding because the discrete exploration {needs} to be constructed such that it discards configurations with not available circuit Hamiltonian, i.e., we need to eliminate frozen and free terms corresponding to nodes where more than one inductor/capacitor are connected~\cite{Chitta2022}. Another consideration relies on the exponential {increase} of the Hilbert space with the number of nodes, which makes such exploration intractable by brute-force methods. 

{To address this problem, we adopt an evolutionary optimization approach following the guidelines presented in Ref.~\cite{Cardenas2023}. Further technical details and numerical convergence analyses can be found in that reference. However, for completeness, we provide a qualitative description here.} The goal of the algorithm is to find a circuit topology $\mathcal{T}$ and its correspondent component values $\mathcal{P}_{\mathcal{T}}$ that best match the requirements established by the cost function $\mathcal{F}$. {The cost function used to tackle the optimization problem consisted of four terms
\begin{eqnarray}
    \text{d}_1 &=& \frac{1}{\min\left(\frac{\abs{\omega_{01}-\omega_{ij}}}{\omega_{01}}{} +\epsilon\right)}\quad \text{for}\quad \begin{cases}
    i\in \{0,1\}\\
    j> i
\end{cases},\\
    \text{d}_2 &=& \abs{\max\left(\abs{\langle 0 | \hat{n}_p | 1 \rangle}-1\right)}\,\,\, \text{for} \,\,\, p\in\{1,\cdots,N-1\},\\
    \text{d}_3 &=& \frac{\abs{\frac{\partial\omega_{01}}{\partial\varphi_\text{ext}}}_{\varphi_\text{ext}=0.5 - \delta\varphi_\text{ext}}}{\xi_{\varphi_\text{ext}}},\\
    \text{d}_4 &=& \frac{\abs{\frac{\partial\omega_{01}}{\partial n_\text{g}}}_{n_\text{g}=0.5 - \delta n_\text{g}}}{\xi_{n_\text{g}}},
\end{eqnarray}
where $\omega_{ij} = \omega_j - \omega_i$ are frequency differences between levels, $\epsilon$ is a correction term to avoid divergence, $\hat{n}_p$ is the charge operator of node $p$, $N$ is the total number of nodes, $\delta\varphi_\text{ext} = 0.01$ and $\delta n_\text{g} = 0.25$ represent external magnetic flux and charge fluctuations, and $\xi_{\varphi_\text{ext}} = 10 \,\text{MHz}$ and $\xi_{n_\text{g}} = 10 \,\text{kHz}$ are normalization factors obtained from reference magnetic flux and charge dispersion values. The total cost function to minimize is given by $\mathcal{F} = \sum_i\text{d}_i^2$.} The circuit structure considered for the optimization is the one depicted in Fig.~\ref{fig:optimization}, where we have established all-to-all connectivity between superconducting islands, assuming always a capacitive coupling, and either a linear and(or) nonlinear inductor connection. The physical values of the circuit components were constrained to fabrication capabilities, to avoid unrealistic capacitive relations or extreme component values, complicating fabrication and reproducibility~\cite{Wedin2017}. The optimization results were furthermore restricted to systems including three modes, discarding configurations expressing a lower number of modes. The recombination process was performed by properly combining suitable modes of the parent configurations, to obtain a resulting circuit with characteristics similar to the previous configurations. Mutations were performed over the circuit components and their physical values, by properly choosing the probability of mutation $p$. The obtained solutions were observed not to represent global minima of the cost function but rather local solutions. While these solutions did not fully meet the cost function requirements, they showed resilience under small deviations in the continuous optimization parameters.

The optimization of the system relies on targeting the low energy sector of the physical device. The algorithm computes the characteristics of the system in terms of node variables. Then takes the lowest energy mode, best matching the characteristics targeted, and combines it with other higher-energy ones. As the algorithm works in terms of node variables, to fully include one mode, it needs to include all circuit components connected to the most optimal node. The low energy modes are usually observed as dispersively coupled to the rest of the system modes, although the large coupling values make this assumption intractable in many cases, for that reason the optimization procedure remains under investigation.

\bibliography{bibRMQ}% Produces the bibliography via BibTeX.

@article{Bouchiat1998,
  title={Quantum coherence with a single Cooper pair},
  author={V. Bouchiat and D. Vion and P. Joyez and D. Esteve and M. H. Devoret},
  journal={Physica Scripta},
  volume={1998},
  number={T76},
  pages={165},
  year={1998},
  doi={10.1238/Physica.Topical.076a00165},
  url={https://doi.org/10.1238/Physica.Topical.076a00165}
}

@article{Nakamura1999,
  author = {Y. Nakamura and Yu. A. Pashkin and J. S. Tsai},
  title = {Coherent control of macroscopic quantum states in a single-Cooper-pair box},
  journal = {Nature},
  year = {1999},
  volume = {398},
  pages = {786-788},
  doi = {10.1038/19718},
  url = {https://www.nature.com/articles/19718}
}

@article{Mooij1999,
  author = {J. E. Mooij and T. P. Orlando and L. Levitov and L. Tian and C. H. Van der Wall and S. Lloyd},
  title = {Josephson Persistent-Current Qubit},
  journal = {Science},
  year = {1999},
  volume = {285},
  number = {5430},
  pages = {1036-1039},
  doi={10.1126/science.285.5430.1036},
  url={https://doi.org/10.1126/science.285.5430.1036}
}

@article{PhysRevA76042319,
  author = {J. Koch and T. M. Yu and J. Gambetta and A. A. Houck and D. I. Schuster and J. Majer and A. Blais and M. H. Devoret and S. M. Girvin and R. J. Schoelkopf},
  title = {Charge-insensitive qubit design derived from the Cooper pair box},
  journal = {Physical Review A},
  year = {2007},
  volume = {76},
  pages = {042319},
  doi={10.1103/PhysRevA.76.042319},
  url={https://doi.org/10.1103/PhysRevA.76.042319}
}

@article{Manucharyan2009,
  author = {V. E. Manucharyan and J. Koch and L. I. Glazman and M. H. Devoret},
  title = {Fluxonium: Single Cooper-Pair Circuit Free of Charge Offsets},
  journal = {Science},
  year = {2009},
  volume = {326},
  number = {5949},
  pages = {113-116},
  doi={10.1126/science.1175552},
  url={https://doi.org/10.1126/science.1175552}
}

@article{Place2021,
  author = {A. P. M. Place and L. V. H. Rodgers and P. Mundada and B. M. Smitham and M. Fitzpatrick and Z. Leng and A. Premkumar and J. Bryon and A. Vrajitoarea and S. Sussman and G. Cheng and T. Madhavan and H. K. Babla and X. H. Le and Y. Gang and B. Jäck and A. Gyenis and N. Yao and R. J. Cava and N. P. de Leon and A. A. Houck},
  title = {New material platform for superconducting transmon qubits with coherence times exceeding 0.3 milliseconds},
  journal = {Nature Communications},
  year = {2021},
  volume = {12},
  pages = {1779},
  doi = {10.1038/s41467-021-22030-5},
  url={https://www.nature.com/articles/s41467-021-22030-5}
}

@article{Wang2022,
  author = {C. Wang and X. Li and H. Xu and Z. Li and J. Wang and Z. Yang and Z. Mi and X. Liang and T. Su and C. Yang and G. Wang and W. Wang and Y. Li and M. Chen and C. Li and K. Linghu and J. Han and Y. Zhang and Y. Feng and Y. Song and T. Ma and J. Zhang and R. Wang and P. Zhao and W. Liu and G. Xue and Y. Jin and H. Yu},
  title = {Towards practical quantum computers: Transmon qubits with long coherence time and high fidelity},
  journal = {npj Quantum Information},
  year = {2022},
  volume = {8},
  pages = {3},
  doi = {10.1038/s41534-021-00510-2},
  url={https://www.nature.com/articles/s41534-021-00510-2}
}

@article{Somoroff2023,
  author = {A. Somoroff and Q. Ficheux and R. A. Mencia and H. Xiong and R. Kuzmin and V. E. Manucharyan},
  title = {Millisecond coherence in a fluxonium qubit with minimal dephasing},
  journal = {Physical Review Letters},
  year = {2023},
  volume = {130},
  pages = {267001},
  doi={10.1103/PhysRevLett.130.267001},
  url={https://doi.org/10.1103/PhysRevLett.130.267001}
}

@article{Gyenis2021,
  author = {A. Gyenis and A. DiPaolo and J. Koch and A. Blais and A. A. Houck and D. I. Schuster},
  title = {Moving beyond the Transmon: Noise-Protected Superconducting Quantum Circuits},
  journal = {PRX Quantum},
  year = {2021},
  volume = {2},
  pages = {030101},
  doi={10.1103/PRXQuantum.2.030101},
  url={https://doi.org/10.1103/PRXQuantum.2.030101}
}

@article{Calzona2023,
  author = {A. Calzona and M. Carrega},
  title = {Multi-mode architectures for noise-resilient superconducting qubits},
  journal = {Superconductor Science and Technology},
  year = {2022},
  volume = {36},
  pages = {023001},
  doi = {10.1088/1361-6668/acaa64},
  url={https://iopscience.iop.org/article/10.1088/1361-6668/acaa64}
}

@article{Ioffe2002,
  author = {L. B. Ioffe and M. V. Feigel'man and A. Ioselevich and D. Ivanov and M. Troyer and G. Blatter},
  title = {Topologically protected quantum bits using Josephson junction arrays},
  journal = {Nature},
  year = {2002},
  volume = {415},
  pages = {503-506},
  doi = {10.1038/415503a},
  url={https://www.nature.com/articles/415503a}
}

@article{Gladchenko2009,
  author = {S. Gladchenko and D. Olaya and E. Dupont-Ferrier and B. Douçot and L. B. Ioffe and M. E. Gershenson},
  title = {Superconducting nanocircuits for topologically protected qubits},
  journal = {Nature Physics},
  year = {2009},
  volume = {5},
  pages = {48-53},
  doi = {10.1038/nphys1151},
  url={https://www.nature.com/articles/nphys1151}
}

@article{Douçot2012,
  author = {B. Douçot and L. B. Ioffe},
  title = {Physical implementation of protected qubits: A review},
  journal = {Reports on Progress in Physics},
  year = {2012},
  volume = {75},
  pages = {072001},
  doi = {10.1088/0034-4885/75/7/072001},
  url={https://iopscience.iop.org/article/10.1088/0034-4885/75/7/072001/meta}
}

@article{Brooks2013,
  author = {P. Brooks and A. Kitaev and J. Preskill},
  title = {Protected gates for superconducting qubits},
  journal = {Physical Review A},
  year = {2013},
  volume = {87},
  pages = {052306},
  doi={10.1103/PhysRevA.87.052306},
  url={https://doi.org/10.1103/PhysRevA.87.052306}
}

@article{Smith2020,
  author = {W. C. Smith and A. Kou and X. Xiao and U. Vool and M. H. Devoret},
  title = {Superconducting circuit protected by two-Cooper-pair tunneling},
  journal = {npj Quantum Information},
  year = {2020},
  volume = {6},
  pages = {8},
  doi = {10.1038/s41534-019-0231-2},
  url={https://www.nature.com/articles/s41534-019-0231-2}
}

@article{Kalashnikov2020,
  author = {K. Kalashnikov and W. T. Hsieh and W. Zhang and W. Lu and P. Kamenov and A. DiPaolo and A. Blais and M. E. Gershenson and M. Bell},
  title = {Bifluxon: Fluxon-Parity-Protected Superconducting Qubit},
  journal = {PRX Quantum},
  year = {2020},
  volume = {1},
  pages = {010307},
  doi={10.1103/PRXQuantum.1.010307},
  url={https://doi.org/10.1103/PRXQuantum.1.010307}
}

@article{Hyyppä2022,
  author = {E. Hyyppä and S. Kundu and C. F. Chan and A. Gunyhó and J. Hotari and D. Janzso and K. Juliusson and O. Kiuru and J. Kotilahti and A. Landra and W. Liu and F. Marxer and A. Mäkinen and J. Orgiazzi and M. Palma and M. Savytskyi and F. Tosto and J. Tuorila and V. Vadimov and T. Li and C. Ockeloen-Korppi and J. Heinsoo and K. Y. Tan and J. Hassel and M. Möttönen},
  title = {Unimon qubit},
  journal = {Nature Communications},
  year = {2022},
  volume = {13},
  pages = {6895},
  doi = {10.1038/s41467-022-34614-w},
  url={https://www.nature.com/articles/s41467-022-34614-w}
}

@article{Mencia2024,
  title = {Integer Fluxonium Qubit},
  author = {Mencia, Raymond A. and Lin, Wei-Ju and Cho, Hyunheung and Vavilov, Maxim G. and Manucharyan, Vladimir E.},
  journal = {PRX Quantum},
  volume = {5},
  issue = {4},
  pages = {040318},
  numpages = {12},
  year = {2024},
  month = {Nov},
  publisher = {American Physical Society},
  doi = {10.1103/PRXQuantum.5.040318},
  url = {https://link.aps.org/doi/10.1103/PRXQuantum.5.040318}
}

@article{Cardenas2023,
  title = {Resilient superconducting-element design with genetic algorithms},
  author = {C\'ardenas-L\'opez, F.A. and Retamal, J.C. and Chen, Xi and Romero, G. and Sanz, M.},
  journal = {Phys. Rev. Appl.},
  volume = {23},
  issue = {5},
  pages = {054068},
  numpages = {20},
  year = {2025},
  month = {May},
  publisher = {American Physical Society},
  doi = {10.1103/PhysRevApplied.23.054068},
  url = {https://link.aps.org/doi/10.1103/PhysRevApplied.23.054068}
}

@article{Leggett1987,
  author = {A. J. Leggett and S. Chakravarty and A. T. Dorsey and M. P. A. Fisher and A. Garg and W. Zwerger},
  title = {Dynamics of the Dissipative Two-State System},
  journal = {Reviews of Modern Physics},
  volume = {67},
  pages = {725},
  year = {1987},
  doi = {10.1103/RevModPhys.67.725},
  url={https://doi.org/10.1103/RevModPhys.67.725}
}

@article{Groszkowski2018,
  author = {P. Groszkowski and A. DiPaolo and A. L. Grimsmo and A. Blais and D. I. Schuster and A. A. Houck and J. Koch},
  title = {Coherence properties of the 0-$\pi$ qubit},
  journal = {New Journal of Physics},
  volume = {20},
  pages = {043053},
  year = {2018},
  doi = {10.1088/1367-2630/aab7cd},
  url={https://doi.org/10.1088/1367-2630/aab7cd}
}

@article{Groszkowski2021,
  author = {P. Groszkowski and J. Koch},
  title = {Scqubits: a Python package for superconducting qubits},
  journal = {Quantum},
  volume = {5},
  pages = {583},
  year = {2021},
  doi = {10.22331/q-2021-09-30-583},
  url={https://doi.org/10.22331/q-2021-09-30-583}
}

@article{Ithier2005,
  author = {G. Ithier and E. Collin and P. Joyez and P. J. Meeson and D. Vion and D. Esteve and F. Chiarello and A. Shnirman and Y. Makhlin and J. Schriefl and G. Schön},
  title = {Decoherence in a superconducting quantum bit circuit},
  journal = {Physical Review B},
  volume = {72},
  pages = {134519},
  year = {2005},
  doi = {10.1103/PhysRevB.72.134519},
  url={https://doi.org/10.1103/PhysRevB.72.134519}
}

@book{Pozar,
  author = {D. M. Pozar},
  title = {Microwave Engineering},
  edition = {4th},
  publisher = {Wiley},
  year = {2012},
  address = {New York}
}

@article{Krantz2019,
  author = {P. Krantz and M. Kjaergaard and F. Yan and T. P. Orlando and S. Gustavsson and W. D. Oliver},
  title = {A Quantum Engineer's Guide to Superconducting Qubits},
  journal = {Applied Physics Reviews},
  volume = {6},
  pages = {021318},
  year = {2019},
  doi = {10.1063/1.5089550},
  url={https://doi.org/10.1063/1.5089550}
}

@article{Nature508,
  author = {I. M. Pop and K. Geerlings and G. Catelani and R. J. Schoelkopf and L. I. Glazman and M. H. Devoret},
  title = {Coherent suppression of electromagnetic dissipation due to superconducting quasiparticles},
  journal = {Nature},
  volume = {508},
  pages = {369-372},
  year = {2014},
  doi = {10.1038/nature13017},
  url={https://doi.org/10.1038/nature13017}
}

@article{PhysRevX9041041,
  author = {L. B. Nguyen and Y. Lin and A. Somoroff and R. Mencia and N. Grabon and V. E. Manucharyan},
  title = {High-Coherence Fluxonium Qubit},
  journal = {Physical Review X},
  volume = {9},
  pages = {041041},
  year = {2019},
  doi = {10.1103/PhysRevX.9.041041},
  url={https://doi.org/10.1103/PhysRevX.9.041041}
}

@article{NelderMead,
  author = {P. van Mulbregt and SciPy 1.0 Contributors},
  title = {SciPy 1.0: Fundamental Algorithms for Scientific Computing in Python},
  journal = {Nature Methods},
  volume = {17},
  pages = {261},
  year = {2020},
  doi = {10.1038/s41592-019-0686-2},
  url={https://doi.org/10.1038/s41592-019-0686-2}
}

@article{PhysRevA69062320,
  author = {A. Blais and R.-S. Huang and A. Wallraff and S. M. Girvin and R. J. Schoelkopf},
  title = {Cavity quantum electrodynamics for superconducting electrical circuits: An architecture for quantum computation},
  journal = {Physical Review A},
  volume = {69},
  pages = {062320},
  year = {2004},
  doi = {10.1103/PhysRevA.69.062320},
  url={https://doi.org/10.1103/PhysRevA.69.062320}
}

@article{PhysRevLett105223601,
  author = {R. Bianchetti and S. Filipp and M. Baur and J. M. Fink and C. Lang and L. Steffen and M. Boissonneault and A. Blais and A. Wallraff},
  title = {Control and Tomography of a Three Level Superconducting Artificial Atom},
  journal = {Physical Review Letters},
  volume = {105},
  pages = {223601},
  year = {2010},
  doi = {10.1103/PhysRevLett.105.223601},
  url={https://doi.org/10.1103/PhysRevLett.105.223601}
}

@article{Sete2015,
  author = {E. A. Sete and J. M. Martinis and A. N. Korotkov},
  title = {Quantum theory of a bandpass Purcell filter for qubit readout},
  journal = {Physical Review A},
  volume = {92},
  pages = {012325},
  year = {2015},
  doi = {10.1103/PhysRevA.92.012325},
  url={https://doi.org/10.1103/PhysRevA.92.012325}
}

@article{Ionization,
  author = {R. Shillito and A. Petrescu and J. Cohen and J. Beall and M. Hauru and M. Ganahl and A. G. M. Lewis and G. Vidal and A. Blais},
  title = {Ionization of a Josephson Junction Artificial Atom},
  journal = {Physical Review Applied},
  volume = {18},
  pages = {034031},
  year = {2022},
  doi = {10.1103/PhysRevApplied.18.034031},
  url={https://doi.org/10.1103/PhysRevApplied.18.034031}
}

@article{PhysRevA77060305,
  author = {M. Boissonneault and J. M. Gambetta and A. Blais},
  title = {Dispersive regime of circuit QED: Photon-dependent qubit dephasing and relaxation rates},
  journal = {Physical Review A},
  volume = {77},
  pages = {060305(R)},
  year = {2008},
  doi = {10.1103/PhysRevA.77.060305},
  url={https://doi.org/10.1103/PhysRevA.77.060305}
}

@article{Motzoi2018,
  author = {F. Motzoi and L. Buchmann and C. Dickel},
  title = {Simple, smooth and fast pulses for dispersive measurements in cavities and quantum networks},
  journal = {arXiv preprint arXiv:1809.04116},
  year = {2018},
  doi = {10.48550/arXiv.1809.04116},
  url = {https://arxiv.org/abs/1809.04116}
}

@article{PhysRevB84064517,
  author = {G. Catelani and R. J. Schoelkopf and M. H. Devoret and L. I. Glazman},
  title = {Relaxation and Frequency Shifts Induced by Quasiparticles in Superconducting Qubits},
  journal = {Physical Review B},
  volume = {84},
  pages = {064517},
  year = {2011},
  doi = {10.1103/PhysRevB.84.064517},
  url={https://doi.org/10.1103/PhysRevB.84.064517}
}

@article{PhysRevLett103217004,
  author = {J. Koch and V. Manucharyan and M. H. Devoret and L. I. Glazman},
  title = {Charging effects in the inductively shunted Josephson junction: Model of the fluxonium qubit},
  journal = {Physical Review Letters},
  volume = {103},
  pages = {217004},
  year = {2009},
  doi = {10.1103/PhysRevLett.103.217004},
  url={https://doi.org/10.1103/PhysRevLett.103.217004}
}

@article{PhysRevX11011010,
  author = {H. Zhang and S. Chakram and T. Roy and N. Earnest and Y. Lu and Z. Huang and D. K. Weiss and J. Koch and D. I. Schuster},
  title = {Universal Fast-Flux Control of a Coherent, Low-Frequency Qubit},
  journal = {Physical Review X},
  volume = {11},
  pages = {011010},
  year = {2021},
  doi = {10.1103/PhysRevX.11.011010},
  url={https://doi.org/10.1103/PhysRevX.11.011010}
}

@article{PRXQuantum3037001,
  author = {L. B. Nguyen and G. Koolstra and Y. Kim and A. Morvan and T. Chistolini and S. Singh and K. N. Nesterov and C. Jünger and L. Chen and Z. Pedramrazi and B. K. Mitchell and J. M. Kreikebaum and S. Puri and D. I. Santiago and I. Siddiqi},
  title = {Blueprint for a High-Performance Fluxonium Quantum Processor},
  journal = {PRX Quantum},
  volume = {3},
  pages = {037001},
  year = {2022},
  doi = {10.1103/PRXQuantum.3.037001},
  url={https://doi.org/10.1103/PRXQuantum.3.037001}
}

@article{PRXQuantum2010339,
  author = {A. Gyenis and P. S. Mundada and A. Di Paolo and T. M. Hazard and X. You and D. I. Schuster and J. Koch and A. Blais and A. A. Houck},
  title = {Experimental Realization of a Protected Superconducting Circuit Derived from the 0–$\pi$ Qubit},
  journal = {PRX Quantum},
  volume = {2},
  pages = {010339},
  year = {2021},
  doi = {10.1103/PRXQuantum.2.010339},
  url={https://doi.org/10.1103/PRXQuantum.2.010339}
}

@article{PhysRevX7031037,
  author = {A. Kou and W. C. Smith and U. Vool and R. T. Brierley and H. Meier and L. Frunzio and S. M. Girvin and L. I. Glazman and M. H. Devoret},
  title = {Fluxonium-based artificial molecules with tunable magnetic interaction between two superconducting qubits},
  journal = {Physical Review X},
  volume = {7},
  pages = {031037},
  year = {2017},
  doi = {10.1103/PhysRevX.7.031037},
  url={https://doi.org/10.1103/PhysRevX.7.031037}
}

@article{PhysRevLett97167001,
  author = {F. Yoshihara and K. Harrabi and A. O. Niskanen and Y. Nakamura and J. S. Tsai},
  title = {Decoherence of Flux Qubits due to 1/f Flux Noise},
  journal = {Physical Review Letters},
  volume = {97},
  pages = {167001},
  year = {2006},
  doi = {10.1103/PhysRevLett.97.167001},
  url={https://doi.org/10.1103/PhysRevLett.97.167001}
}

@article{Pedersen2007,
  author = {L. H. Pedersen and N. M. Møller and K. Mølmer},
  title = {Fidelity of quantum operations},
  journal = {Physics Letters A},
  volume = {367},
  pages = {47-51},
  year = {2007},
  doi = {10.1016/j.physleta.2007.02.069},
  url={https://doi.org/10.1016/j.physleta.2007.02.069}
}

@article{Motzoi2009,
  author = {F. Motzoi and J. M. Gambetta and P. Rebentrost and F. K. Wilhelm},
  title = {Simple Pulses for Elimination of Leakage in Weakly Nonlinear Qubits},
  journal = {Physical Review Letters},
  volume = {103},
  pages = {110501},
  year = {2009},
  doi = {10.1103/PhysRevLett.103.110501},
  url={https://doi.org/10.1103/PhysRevLett.103.110501}
}

@article{Chen2016,
  author = {Z. Chen and, J. Kelly and C. Quintana and R. Barends and B. Campbell and Y. Chen and B. Chiaro and A. Dunsworth and A. G. Fowler and E. Lucero and E. Jeffrey and A. Megrant and J. Mutus and M. Neeley and C. Neill and P. J. J. O’Malley and P. Roushan and D. Sank and A. Vainsencher and J. Wenner and T. C. White and A. N. Korotkov and John M. Martinis},
  title = {Measuring and Suppressing Quantum State Leakage in a Superconducting Qubit},
  journal = {Physical Review Letters},
  volume = {116},
  pages = {020501},
  year = {2016},
  doi = {10.1103/PhysRevLett.116.020501},
  url={https://doi.org/10.1103/PhysRevLett.116.020501}
}

@article{Motzoi2013,
  author = {F. Motzoi and F. K. Wilhelm},
  title = {Improving the gate fidelity of encoded qubits by optimal control of decoupling pulses},
  journal = {Physical Review A},
  volume = {88},
  pages = {062318},
  year = {2013},
  doi = {10.1103/PhysRevA.88.062318},
  url={https://doi.org/10.1103/PhysRevA.88.062318}
}

@article{Pfeiffer2023,
  author = {Frederik Pfeiffer and Max Werninghaus and Christian Schweizer and Niklas Bruckmoser and Leon Koch and Niklas J. Glaser and Gerhard Huber and David Bunch and Franz X. Haslbeck and M. Knudsen and Gleb Krylov and Klaus Liegener and Achim Marx and Lea Richard and João H. Romeiro and Federico Roy and Johannes Schirk and Christian Schneider and Malay Singh and Lasse Södergren and Ivan Tsitsilin and Florian Wallner and Carlos A. Riofrío and Stefan Filipp},
  title = {Efficient decoupling of a non-linear qubit mode from its environment},
  journal = {Phys. Rev. X},
  volume = {14},
  issue = {4},
  pages = {041007},
  numpages = {18},
  year = {2024},
  month = {Oct},
  publisher = {American Physical Society},
  doi = {10.1103/PhysRevX.14.041007},
  url = {https://link.aps.org/doi/10.1103/PhysRevX.14.041007}
}

@manual{Fastcap1,
  author = {S. R. L.},
  title = {FastFieldSolvers},
  note = {Fast Field Solvers products}
}

@article{Fastcap2,
  author = {K. Nabors and J. White},
  title = {FastCap: A multipole accelerated 3-D capacitance extraction program},
  journal = {IEEE Transactions on Computer-Aided Design of Integrated Circuits and Systems},
  volume = {10},
  pages = {1447-1459},
  year = {1991},
  doi = {10.1109/43.97611},
  url={https://doi.org/10.1109/43.97611}
}

@article{Chitta2022,
  author = {S. P. Chitta and T. Zhao and Z. Huang and I. Mondragon-Shem and J. Koch},
  title = {Computer-aided quantization and numerical analysis of superconducting circuits},
  journal = {New Journal of Physics},
  volume = {24},
  pages = {103020},
  year = {2022},
  doi = {10.1088/1367-2630/ac8d5c},
  url={https://doi.org/10.1088/1367-2630/ac8d5c}
}

@article{Wedin2017,
  author = {G. Wedin},
  title = {Quantum information processing with superconducting circuits: a review},
  journal = {Reports on Progress in Physics},
  volume = {80},
  pages = {106001},
  year = {2017},
  doi = {10.1088/1361-6633/aa70e8},
  url={https://doi.org/10.1088/1361-6633/aa70e8}
}

@article{Ganjam2024,
	author = {Ganjam, Suhas and Wang, Yanhao and Lu, Yao and Banerjee, Archan and Lei, Chan U. and Krayzman, Lev and Kisslinger, Kim and Zhou, Chenyu and Li, Ruoshui and Jia, Yichen and Liu, Mingzhao and Frunzio, Luigi and Schoelkopf, Robert J.},
	date = {2024/05/01},
	doi = {10.1038/s41467-024-47857-6},
	id = {Ganjam2024},
	isbn = {2041-1723},
	journal = {Nature Communications},
	number = {1},
	pages = {3687},
	title = {Surpassing millisecond coherence in on chip superconducting quantum memories by optimizing materials and circuit design},
	url = {https://doi.org/10.1038/s41467-024-47857-6},
	volume = {15},
	year = {2024}}

@article{Paik2011,
  title = {Observation of High Coherence in Josephson Junction Qubits Measured in a Three-Dimensional Circuit QED Architecture},
  author = {Paik, Hanhee and Schuster, D. I. and Bishop, Lev S. and Kirchmair, G. and Catelani, G. and Sears, A. P. and Johnson, B. R. and Reagor, M. J. and Frunzio, L. and Glazman, L. I. and Girvin, S. M. and Devoret, M. H. and Schoelkopf, R. J.},
  journal = {Phys. Rev. Lett.},
  volume = {107},
  issue = {24},
  pages = {240501},
  numpages = {5},
  year = {2011},
  month = {Dec},
  publisher = {American Physical Society},
  doi = {10.1103/PhysRevLett.107.240501},
  url = {https://link.aps.org/doi/10.1103/PhysRevLett.107.240501}
}

@article{jerger2024dispersive,
  title={Dispersive qubit readout with intrinsic resonator reset},
  author={Jerger, M and Motzoi, F and Gao, Y and Dickel, C and Buchmann, L and Bengtsson, A and Tancredi, G and Warren, CW and Bylander, J and DiVincenzo, D and others},
  journal={arXiv preprint arXiv:2406.04891},
  year={2024}
}

@article{Egusquiza2022,
  title = {Role of anomalous symmetry in $0\text{\ensuremath{-}}\ensuremath{\pi}$ qubits},
  author = {Egusquiza, I. L. and I\~niguez, A. and Rico, E. and Villarino, A.},
  journal = {Phys. Rev. B},
  volume = {105},
  issue = {20},
  pages = {L201104},
  numpages = {5},
  year = {2022},
  month = {May},
  publisher = {American Physical Society},
  doi = {10.1103/PhysRevB.105.L201104},
  url = {https://link.aps.org/doi/10.1103/PhysRevB.105.L201104}
}

@article{Hyyppa2024,
  title = {Reducing Leakage of Single-Qubit Gates for Superconducting Quantum Processors Using Analytical Control Pulse Envelopes},
  author = {Hyypp\"a, Eric and Veps\"al\"ainen, Antti and Papi\ifmmode \check{c}\else \v{c}\fi{}, Miha and Chan, Chun Fai and Inel, Sinan and Landra, Alessandro and Liu, Wei and Luus, J\"urgen and Marxer, Fabian and Ockeloen-Korppi, Caspar and Orbell, Sebastian and Tarasinski, Brian and Heinsoo, Johannes},
  journal = {PRX Quantum},
  volume = {5},
  issue = {3},
  pages = {030353},
  numpages = {28},
  year = {2024},
  month = {Sep},
  publisher = {American Physical Society},
  doi = {10.1103/PRXQuantum.5.030353},
  url = {https://link.aps.org/doi/10.1103/PRXQuantum.5.030353}
}

@article{Rower2024,
  title = {Suppressing Counter-Rotating Errors for Fast Single-Qubit Gates with Fluxonium},
  author = {Rower, David A. and Ding, Leon and Zhang, Helin and Hays, Max and An, Junyoung and Harrington, Patrick M. and Rosen, Ilan T. and Gertler, Jeffrey M. and Hazard, Thomas M. and Niedzielski, Bethany M. and Schwartz, Mollie E. and Gustavsson, Simon and Serniak, Kyle and Grover, Jeffrey A. and Oliver, William D.},
  journal = {PRX Quantum},
  volume = {5},
  issue = {4},
  pages = {040342},
  numpages = {17},
  year = {2024},
  month = {Dec},
  publisher = {American Physical Society},
  doi = {10.1103/PRXQuantum.5.040342},
  url = {https://link.aps.org/doi/10.1103/PRXQuantum.5.040342}
}

@article{Cohen2023,
  title = {Reminiscence of Classical Chaos in Driven Transmons},
  author = {Cohen, Joachim and Petrescu, Alexandru and Shillito, Ross and Blais, Alexandre},
  journal = {PRX Quantum},
  volume = {4},
  issue = {2},
  pages = {020312},
  numpages = {27},
  year = {2023},
  month = {Apr},
  publisher = {American Physical Society},
  doi = {10.1103/PRXQuantum.4.020312},
  url = {https://link.aps.org/doi/10.1103/PRXQuantum.4.020312}
}

@article{Dumas2024,
  title = {Measurement-Induced Transmon Ionization},
  author = {Dumas, Marie Fr\'ed\'erique and Groleau-Par\'e, Benjamin and McDonald, Alexander and Mu\~noz-Arias, Manuel H. and Lled\'o, Crist\'obal and D'Anjou, Benjamin and Blais, Alexandre},
  journal = {Phys. Rev. X},
  volume = {14},
  issue = {4},
  pages = {041023},
  numpages = {31},
  year = {2024},
  month = {Oct},
  publisher = {American Physical Society},
  doi = {10.1103/PhysRevX.14.041023},
  url = {https://link.aps.org/doi/10.1103/PhysRevX.14.041023}
}

@article{PhysRevA.74.042318,
  title = {Qubit-photon interactions in a cavity: Measurement-induced dephasing and number splitting},
  author = {Gambetta, Jay and Blais, Alexandre and Schuster, D. I. and Wallraff, A. and Frunzio, L. and Majer, J. and Devoret, M. H. and Girvin, S. M. and Schoelkopf, R. J.},
  journal = {Phys. Rev. A},
  volume = {74},
  issue = {4},
  pages = {042318},
  numpages = {14},
  year = {2006},
  month = {Oct},
  publisher = {American Physical Society},
  doi = {10.1103/PhysRevA.74.042318},
  url = {https://link.aps.org/doi/10.1103/PhysRevA.74.042318}
}

@misc{Ateshian2025,
      title={Temperature and Magnetic-Field Dependence of Energy Relaxation in a Fluxonium Qubit}, 
      author={Lamia Ateshian and Max Hays and David A. Rower and Helin Zhang and Kate Azar and R\'eouven Assouly and Leon Ding and Michael Gingras and Hannah Stickler and Bethany M. Niedzielski and Mollie E. Schwartz and Terry P. Orlando and Joel \^I-j. Wang and Simon Gustavsson and Jeffrey A. Grover and Kyle Serniak and William D. Oliver},
      year={2025},
      eprint={2507.01175},
      archivePrefix={arXiv},
      primaryClass={quant-ph},
      url={https://arxiv.org/abs/2507.01175}, 
}

@misc{weissler2024,
      title={Enumeration of all superconducting circuits up to 5 nodes}, 
      author={Eli J. Weissler and Mohit Bhat and Zhenxing Liu and Joshua Combes},
      year={2024},
      eprint={2410.18497},
      archivePrefix={arXiv},
      primaryClass={quant-ph},
      url={https://arxiv.org/abs/2410.18497}, 
}

@article{Houck2008,
  title = {Controlling the Spontaneous Emission of a Superconducting Transmon Qubit},
  author = {Houck, A. A. and Schreier, J. A. and Johnson, B. R. and Chow, J. M. and Koch, Jens and Gambetta, J. M. and Schuster, D. I. and Frunzio, L. and Devoret, M. H. and Girvin, S. M. and Schoelkopf, R. J.},
  journal = {Phys. Rev. Lett.},
  volume = {101},
  issue = {8},
  pages = {080502},
  numpages = {4},
  year = {2008},
  month = {Aug},
  publisher = {American Physical Society},
  doi = {10.1103/PhysRevLett.101.080502},
  url = {https://link.aps.org/doi/10.1103/PhysRevLett.101.080502}
}

@article{Dogan2023,
  title = {Two-Fluxonium Cross-Resonance Gate},
  author = {Dogan, Ebru and Rosenstock, Dario and Le Guevel, Lo\"{\i}ck and Xiong, Haonan and Mencia, Raymond A. and Somoroff, Aaron and Nesterov, Konstantin N. and Vavilov, Maxim G. and Manucharyan, Vladimir E. and Wang, Chen},
  journal = {Phys. Rev. Appl.},
  volume = {20},
  issue = {2},
  pages = {024011},
  numpages = {19},
  year = {2023},
  month = {Aug},
  publisher = {American Physical Society},
  doi = {10.1103/PhysRevApplied.20.024011},
  url = {https://link.aps.org/doi/10.1103/PhysRevApplied.20.024011}
}

\end{document}